\documentclass[aps,prd,nofootinbib,twocolumn,superscriptaddress,preprintnumbers]{revtex4}
\pdfoutput=1

\usepackage{amssymb}
\usepackage{amsmath}
\usepackage{epsfig}
\usepackage{breakurl}
\usepackage{comment}

\makeatletter
\def\simgt{\mathrel{\lower2.5pt\vbox{\lineskip=0pt\baselineskip=0pt
           \hbox{$>$}\hbox{$\sim$}}}}
\def\simlt{\mathrel{\lower2.5pt\vbox{\lineskip=0pt\baselineskip=0pt
           \hbox{$<$}\hbox{$\sim$}}}}
\makeatother

\newcommand{\be}{\begin{equation}}
\newcommand{\ee}{\end{equation}}
\newcommand{\bea}{\begin{eqnarray}}
\newcommand{\eea}{\end{eqnarray}}
\newcommand{\Fig}[1]{Fig.~\ref{#1}}

\newcommand{\Eq}[1]{Eq.~(\ref{#1})}

\newcommand{\Sec}[1]{Sec.~\ref{#1}}

\newcommand{\mPl}{m_{\rm Pl}}

\begin{document}

\title{Bubble Baryogenesis}

\author{Clifford Cheung}
\affiliation{Berkeley Center for Theoretical Physics, 
  University of California, Berkeley, CA 94720, USA}
\affiliation{Theoretical Physics Group, 
  Lawrence Berkeley National Laboratory, Berkeley, CA 94720, USA}

\author{Alex Dahlen}
\affiliation{Berkeley Center for Theoretical Physics, 
  University of California, Berkeley, CA 94720, USA}
\affiliation{Theoretical Physics Group, 
  Lawrence Berkeley National Laboratory, Berkeley, CA 94720, USA}

\author{Gilly Elor}
\affiliation{Berkeley Center for Theoretical Physics, 
  University of California, Berkeley, CA 94720, USA}
\affiliation{Theoretical Physics Group, 
  Lawrence Berkeley National Laboratory, Berkeley, CA 94720, USA}

\begin{abstract}
We propose an alternative mechanism of baryogenesis in which a scalar baryon undergoes a percolating first-order phase transition in the early Universe.  The potential barrier that divides the phases contains explicit $B$ and $CP$ violation and the corresponding instanton that mediates decay is therefore asymmetric.  The nucleation and growth of these asymmetric bubbles dynamically generates baryons, which thermalize after percolation; bubble collision dynamics can also add to the asymmetry yield.  We present an explicit toy model that undergoes bubble baryogenesis, and numerically study the evolution of the baryon asymmetry through bubble nucleation and growth, bubble collisions, and washout.  We discuss more realistic constructions, in which the scalar baryon and its potential arise amongst the color-breaking minima of the MSSM, or in the supersymmetric neutrino seesaw mechanism.  Phenomenological consequences, such as gravitational waves, and possible applications to asymmetric dark-matter generation are also discussed.
\end{abstract}


\maketitle

\section{Introduction}

The standard model is incomplete: it does not accommodate the observed baryon asymmetry and therefore new physics is required.  Substantial effort has been devoted to constructing theories that dynamically generate this asymmetry and some prominent contenders include GUT baryogenesis \cite{GUTbaryo}, electroweak baryogenesis \cite{EWbaryo},  thermal leptogenesis \cite{Leptogenesis}, and Affleck-Dine baryogenesis \cite{AD, ADReview}.  This paper proposes a new mechanism, which we dub `bubble baryogenesis'.  

Like the Affleck-Dine mechanism, our setup employs a complex scalar baryon $\phi$, represented in a polar decomposition as
\bea 
\phi(x) &=& R(x) e^{i \theta(x)},
\label{polar}
\eea
where $R(x)$ and $\theta(x)$ are four-dimensional real scalar fields.  Under baryon-number transformations $U(1)_B$, $\phi$ rephases, $R$ is invariant, and $\theta$ shifts.
The charge density of $\phi$ is identified with the number density of baryons
\bea
n_B\equiv {\rm Im} ( \phi^* \dot\phi)=R^2\dot\theta,
\label{eq:B}
\eea
so a baryon asymmetry is present in field configurations that have `angular momentum' in field space.
Constraints on $B$ violation today imply that $\phi$ is currently at the origin of field space---so as not to spontaneously break $B$---and that the potential there has approximate $U(1)_B$---so as not to explicitly break it.  In the early Universe, however, we take $\phi$ to be displaced from this minimum, to a place in the potential where $B$ violation is more substantial.  The observed baryon asymmetry is dynamically generated during the field's journey towards the origin.

In the Affleck-Dine mechanism $\phi$ evolves  \emph{classically}, relaxing uniformly towards the $B$-symmetric minimum.  $B$-violating potential terms torque $\phi$ during its evolution, and so instead of moving in a straight line through field space, $\phi$ takes a curved trajectory; $\phi$ develops non-zero $\dot{\theta}$ and consequently non-zero $B$.  The phase transition from the $B$-violating vacuum in the past to the $B$-symmetric vacuum today is \emph{second-order} or higher-order, and the end result is a \emph{spatially homogeneous} condensate carrying a non-zero baryon asymmetry.

But what if there is no classical trajectory connecting $\phi$ to the symmetric minimum?  Bubble baryogenesis occurs when $\phi$ evolves via \emph{bubble nucleation}---either through quantum tunneling or thermal excitation.  Spherical bubbles of true, $B$-symmetric vacuum nucleate inside the false $B$-violating background.  The bubbles expand, collide, and eventually percolate; the phase transition completes when the entire Universe is in the $B$-symmetric phase.   During this process, baryons are produced through two distinct mechanisms.   First, just as $\phi$ receives a torque in Affleck-Dine, the instanton that mediates bubble nucleation also receives a torque from $B$-violating interactions.  Consequently, the bubble wall takes a curved trajectory through field space, and it therefore accumulates $B$ as it expands.  Second, when the bubble walls collide,  $\phi$ can be excited back into a region of the potential where $B$-violating terms are large, generating additional baryon asymmetry. In bubble baryogenesis, the phase transition is \emph{first-order}, and the end result is a \emph{spatially inhomogeneous} distribution of baryons.  After percolation, the baryon asymmetry is assimilated into the thermal plasma of the early Universe.

Like bubble baryogenesis, electroweak baryogenesis also relies on a first-order phase transition in the early Universe.  In that case, however, the tunneling scalar is the Higgs field, and baryon number is generated indirectly through scattering off bubble walls.

In \Sec{general}, we outline the basic elements of bubble baryogenesis and present a general analysis of the vacuum structure, nucleation rate, asymmetry generation, bubble collisions, and washout.   In \Sec{sec:toy}, we define an explicit toy model, outline its cosmological history, and ascertain the final baryon asymmetry.  Some more realistic examples---involving the neutrino seesaw mechanism, and color breaking minima---are then presented in \Sec{sec:realistic}.  We discuss phenomenological signatures in \Sec{sec:pheno} and conclude in \Sec{sec:conclusions}.

\begin{figure}[t]
\centering
\includegraphics[scale=0.83]{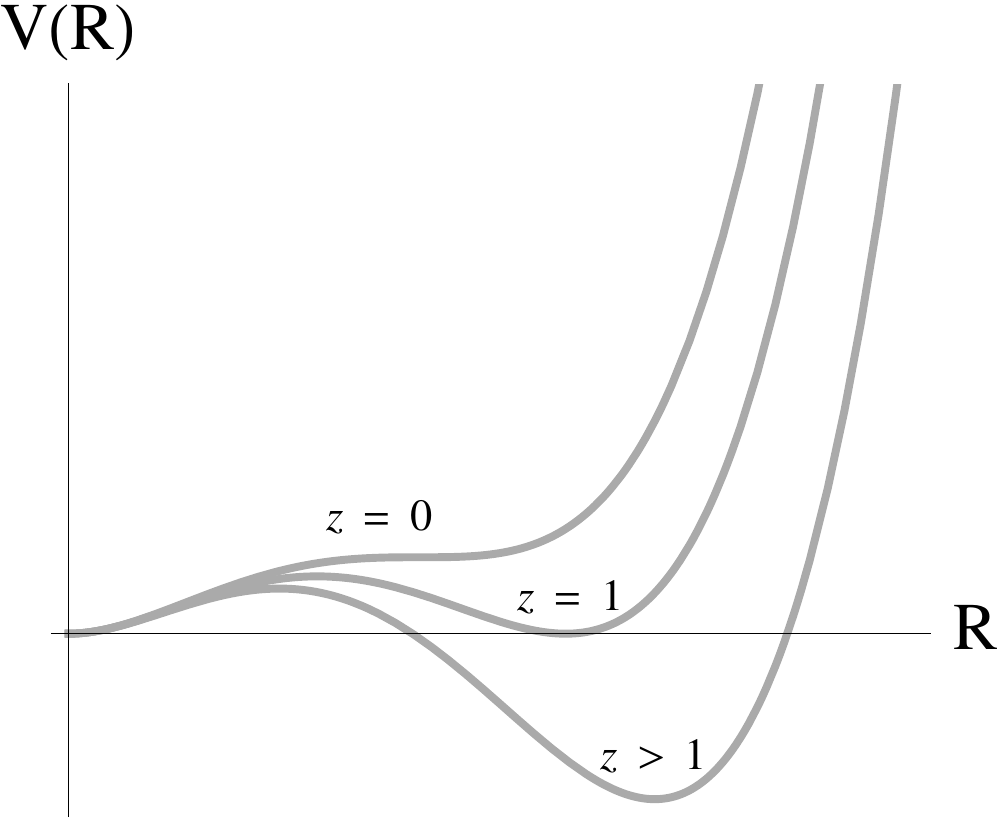} \\
\caption{\footnotesize{A potential that evolves with time so as to give a percolating first-order phase transition; the radial part of our potential has this behavior.  We define a `clock' parameter $z$ that indexes the shape of the potential.  At early times, $z >1$ and the true minimum is $B$-violating and at $\phi\neq0$. At late times, $z<0$ and the only minimum is  $B$-symmetric and at $\phi=0$.  Vacuum decay is possible in the interval $1 > z> 0$.}}
\label{fig:SymV}
\end{figure}

\section{General Considerations}
\label{general}

Of the potential for the scalar baryon $V(\phi)$, we require two features: first, that it vary with time in such a way as to yield a percolating first-order phase transition in the early Universe; and second, that the potential accommodate explicit $B$- and $CP$-breaking dynamics.
We discuss how these two criteria can be accommodated in \Sec{subsec:vacuum} and \Sec{subsec:BCP}, respectively.   Afterwards, we tackle the dynamics of baryon production, which occurs at two times: first, at nucleation, which we discuss in \Sec{subsec:asymmetryNuc}; and second, at collision, which we discuss in \Sec{subsec:asymmetryColl}.  Once generated, the baryon asymmetry must persist, and migrate into the standard-model sector; we discuss the washout and decay of $\phi$ particles in \Sec{sec:washout}.

\subsection{Vacuum Structure and Tunneling}
\label{subsec:vacuum}

A potential that achieves a first-order phase transition must evolve with time as in \Fig{fig:SymV}.  In the early Universe, the stable minimum is $B$-violating and at $\phi\neq0$.  As time evolves, the energy density of this $B$-breaking vacuum increases until it is no longer the true vacuum; the lowest energy vacuum is now $B$-symmetric and at $\phi=0$, and bubble nucleation begins.  At even later times, the energy density in the $B$-breaking vacuum increases so much that the minimum disappears entirely, ensuring that no region of the Universe remains stuck there.  For the transition to be first-order, the bubbles must percolate before this time.

\break

It is convenient to introduce a dimensionless `clock' parameter $z$ that characterizes the evolution of the potential.  The two most important events---when the minima become degenerate and when the $B$-breaking minimum disappears---are taken to be at $z=1$ and $z=0$, respectively, as in \Fig{fig:SymV}.  So,
\bea
\begin{array}{rclll}
&z &\geq 1  & \quad \textrm{stable vacuum at } \phi \neq 0\\
1 > &z& > 0 & \quad  \textrm{metastable vacuum at } \phi\neq 0 \\
0 \geq &z  && \quad \textrm{no vacuum at } \phi \neq 0.
\end{array} 
\label{phases}
\eea
In bubble baryogenesis, $z$ decreases monotonically; it is a reparameterization of time.

Bubble nucleation occurs in the window $1 > z >0$.  Tunneling is mediated by an instanton, a solution to the Euclidean equations of motion with a single negative mode. The rate per unit volume per unit time is given by
\bea 
\Gamma(z)  &=& K(z) e^{-\Delta S(z)},
\label{decayrate}
\eea
where $K$ is a determinant factor of order the characteristic mass scale of the potential to the fourth power, and $ \Delta S$ is the difference in Euclidean action between the instanton and the false vacuum \cite{Coleman}.  At $z=1$, tunneling is forbidden, $\Delta S$ is divergent, and $\Gamma$ is zero.  At $z=0$, $\Delta S$ is zero and semiclassical tunneling through the barrier is overcome by classical evolution down the potential.

After nucleation, the energy difference across the bubble wall causes it to accelerate outward, rapidly approaching the speed of light.  Bubbles nucleate, expand, and collide until---at percolation---the entire Universe is in the true vacuum.  When does percolation occur?

Pick a point in space at time $z$.  The expected number of bubbles $N$ that have overlapped this point is
\bea
N(z)\sim\int_1^{z} V(z,z') \Gamma(z') \frac{dt}{dz'} dz',
\label{percolationcondition}
\eea
where $V(z,z')$ is the three-volume at time $z'$ of the past lightcone that emerged from our point in space at time $z$, and $dt/dz'$ is a Jacobian factor.   The integrand therefore is the probability that a bubble nucleates at a time $z$ in the right position to convert our point to the true vacuum, and it is integrated over all past $z$.  Percolation occurs at a time $z_*$ satisfying
\bea
1\sim N(z_*),
\label{percApprox}
\eea
so that at least one bubble has nucleated in the past lightcone of every point in space.

\begin{figure*}[t]
\centering
\includegraphics[width=7 in]{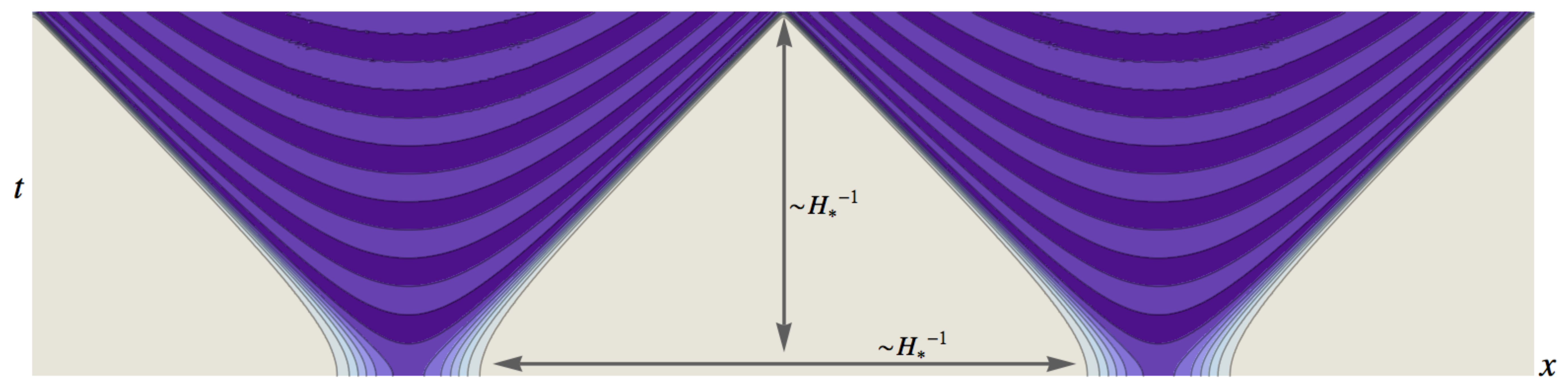} \\
\caption{\footnotesize{Depiction of a simplified bubble collision geometry.  At percolation, there is typically one bubble per Hubble volume and the bubbles nucleated one Hubble time earlier.  Since the surface area of a bubble scales as $H_*^{-2}$ and the volume scales as $H_*^{-3}$, the baryon number density scales as $H_*$.
}}
\label{fig:WallCartoon}
\end{figure*}

 As long as $z$ is changing slowly enough with time, then $N(z_*)\sim\Gamma_* H_*^{-4}$, where throughout the $*$ subscript will denote a quantity evaluated at $z=z_*$.  The integral is dominated by values of $z$ close to $z_*$ and the percolation condition can be approximated by 
 \bea
 \Gamma_*\sim H_*^{\,4}.
 \eea
 \break

\noindent This means that bubbles typically nucleate one Hubble time before percolation, with roughly one bubble per Hubble volume at percolation, as shown in \Fig{fig:WallCartoon}.  Though some bubbles do  nucleate before this time, the rate is too small to induce percolation.

A time-varying potential of the form in \Fig{fig:SymV} can arise naturally in two ways, depending on whether percolation completes before or after reheating.  If the phase transition occurs before reheating, then a direct coupling of the scalar baryon $\phi$ to the inflaton field will give rise to a time-dependent effective potential.  This is the same type of coupling that is used to generate the evolving potential in the Affleck-Dine mechanism \cite{Randall}.  From an effective-field-theory standpoint, such couplings are mandatory unless forbidden by a symmetry, and while they are often non-renormalizable they can nonetheless play an essential role in the physics.  In this scenario, the phase transition takes place between the end of inflation---so as not to dilute the baryons---and reheating.  During this matter-dominated period the inflaton is oscillating about its minimum, but has not yet decayed to standard model particles.

If the phase transition occurs after reheating, then a direct coupling of $\phi$ to the big bang plasma will give rise to a time-dependent thermal correction to the effective potential.  The same couplings that allow $\phi$ to decay---so that the baryon asymmetry can migrate to the standard model sector---can generate such terms.
Note that similar thermal effects give rise to the first-order phase transition in electroweak baryogenesis \cite{hall}, with the crucial difference that the relevant scalar field there, the Higgs boson, is not charged under $U(1)_B$.  In electroweak baryogenesis the purpose of the first-order phase transition is merely to provide an out of equilibrium environment for particle and anti-particle scattering processes.

While bubble baryogenesis can occur in either scenario, the models we study in \Sec{sec:toy} and \Sec{sec:realistic} are in the former category, where it is easier to suppress thermal washout.

\subsection{$B$ and $CP$ Violation}
\label{subsec:BCP}

The Sakharov conditions \cite{sakharov} state that successful baryogenesis requires both $B$- and $CP$-violating dynamics.  
Under $B$ and $CP$ transformations the angular field component transforms as
\bea
\label{eq:Btrans}
\theta &\overset{ B}{\longrightarrow}& \theta + {\rm const} \\
\theta &\overset{CP}{\longrightarrow}& -\theta .
\label{eq:CPtrans}
\eea
By \Eq{eq:Btrans}, $B$ violation requires a potential that violates the shift symmetry on $\theta$, i.e.~carries explicit dependence on $\theta$.    Such terms are necessary for asymmetry generation because in their absence the field has no reason to move in the $\theta$-direction of field space, so by \Eq{eq:B} no asymmetry is produced.  In the Affleck-Dine mechanism, these $B$-violating terms torque the field in the $\theta$-direction on its journey back to the origin; in bubble baryogenesis, they force the instanton, which solves the Euclidean equations of motion, to arc in the $\theta$-direction as a function of spacetime.

By \Eq{eq:CPtrans}, $CP$-violation requires either potential couplings with complex phases or spontaneous breaking by an initial $\phi$ localized at a $CP$-odd minimum.  $CP$ violation is necessary for asymmetry generation because in its absence, though the potential can exert a torque, $\phi$'s trajectory is just as likely to curve in the $+\theta$-direction as it is to curve in the $-\theta$-direction.  That is, in a $CP$-conserving theory, two $CP$-conjugate instantons contribute equally to the path integral. The percolating transition would therefore be comprised of an equal number of bubbles with positive and negative $B$, which average out to a $B$-symmetric Universe.  

Explicit $CP$ violation breaks the degeneracy between these two $CP$-conjugate instantons.   For example, one of them may disappear entirely if it is no longer a saddle point of the action.  Alternatively, both $CP$-conjugate instantons can persist, but the one with a larger associated Euclidean action will be exponentially subdominant to the process of vacuum decay.  This will be true in the models we consider here, so we will only be concerned with the dynamics of the dominant instanton contribution.

\begin{figure}[t]
\centering
\begin{tabular}{c}
\includegraphics[scale=0.45]{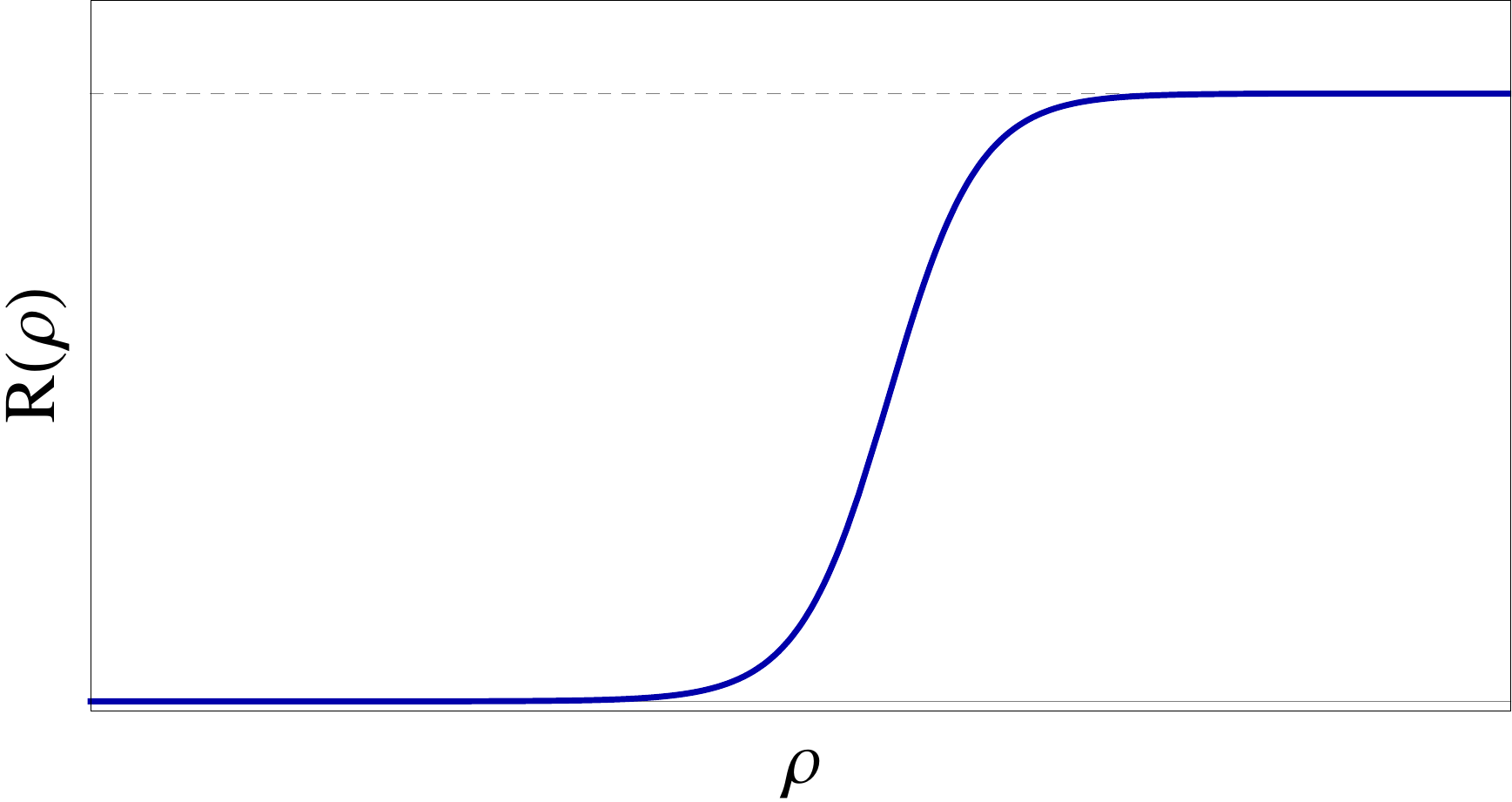}\\
\includegraphics[scale=0.45]{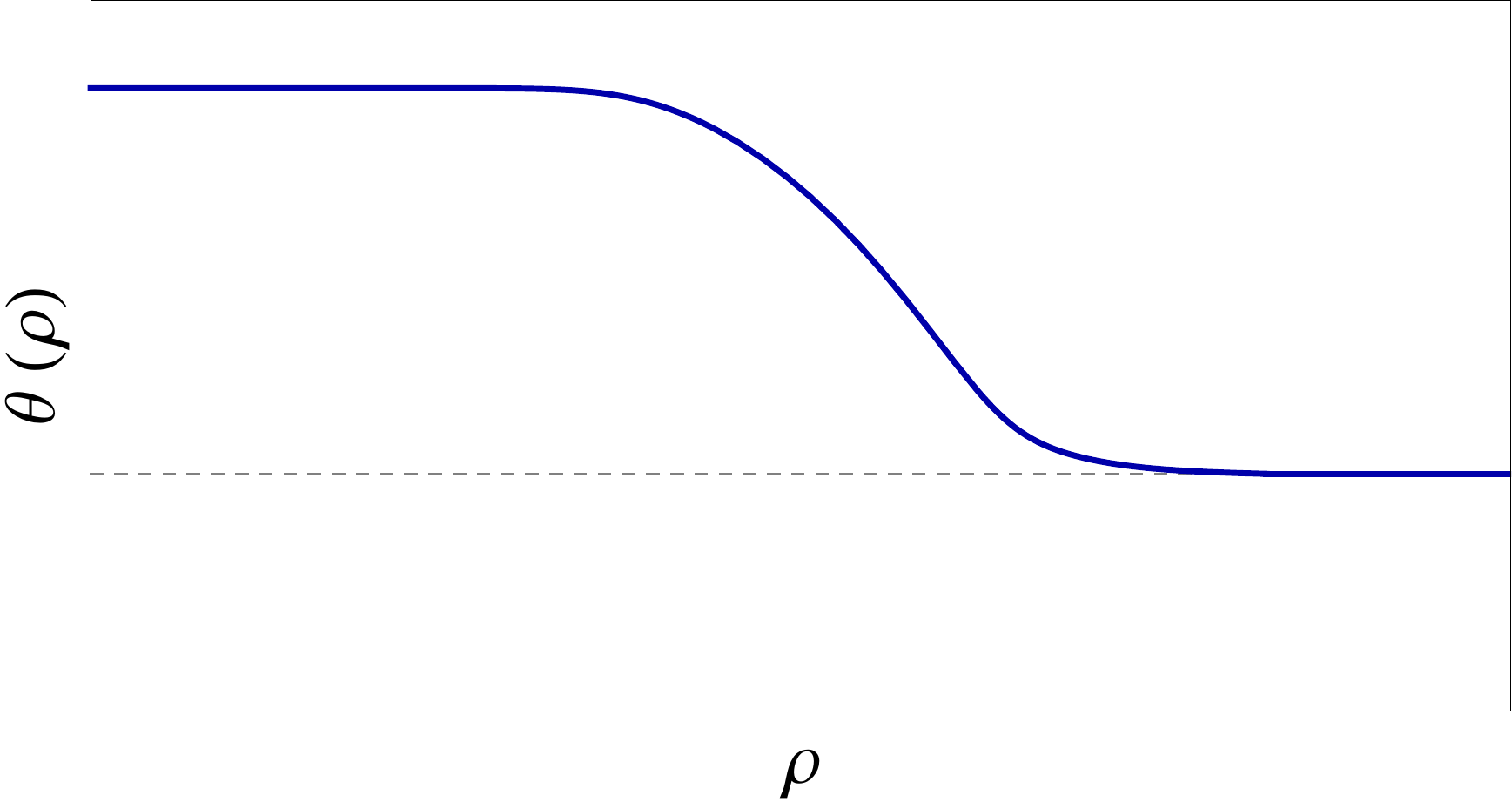}
\end{tabular}
\caption{\footnotesize{Sample instanton profiles for the fields $R$ and $\theta$ as a function of radial coordinate $\rho$ in a potential that breaks $U(1)_B$. False vacuum values for the fields are indicated with dashed gray lines; the true vacuum is at $R=0$. The center of the bubble ($\rho = 0$) is on the left.}}
\label{fig:prof}
\end{figure}

In general, it is useful to characterize the degree of $B$- and $CP$-violating effects with a dimensionless `efficiency' parameter $\epsilon$ which is proportional to $B$- and $CP$-violating parameters in such a way that  $n_B\propto \epsilon$.  From an effective-field-theory perspective,  $\epsilon \ll 1$ is technically natural, but $\epsilon \simeq 1$ is also allowed.


Bubble baryogenesis generates baryon asymmetry in two ways.  First, the instanton itself is asymmetric, which manifests itself as a surface density of baryons on the bubble walls.  Second, bubble collisions excite the field back into the $B$-violating region of the potential.
The net number density of baryons is given by a sum
\bea
\label{eq:totalnb}
n_{B} &=& n_{B,\textrm{instanton}}+n_{B,\textrm{collision}}.
\eea
We will discuss the two contributions in detail in \Sec{subsec:asymmetryNuc} and \Sec{subsec:asymmetryColl}, respectively, and show that, for a broad class of models, both of these contributions scale as
\bea
\label{eq:totalnB}
n_{B,\textrm{instanton}} \sim n_{B,\textrm{collision}} \sim
\epsilon R_\text{F}^{\,2} H_*,
\eea
 where $\epsilon$ is the dimensionless measure of 
$B$ and $CP$ violation, $R_\text{F}= |\phi_\text{F}|$ in the false vacuum, and  $ H_*$ is Hubble at the time of percolation.

\begin{figure}[t]
\centering
\includegraphics[scale=0.45]{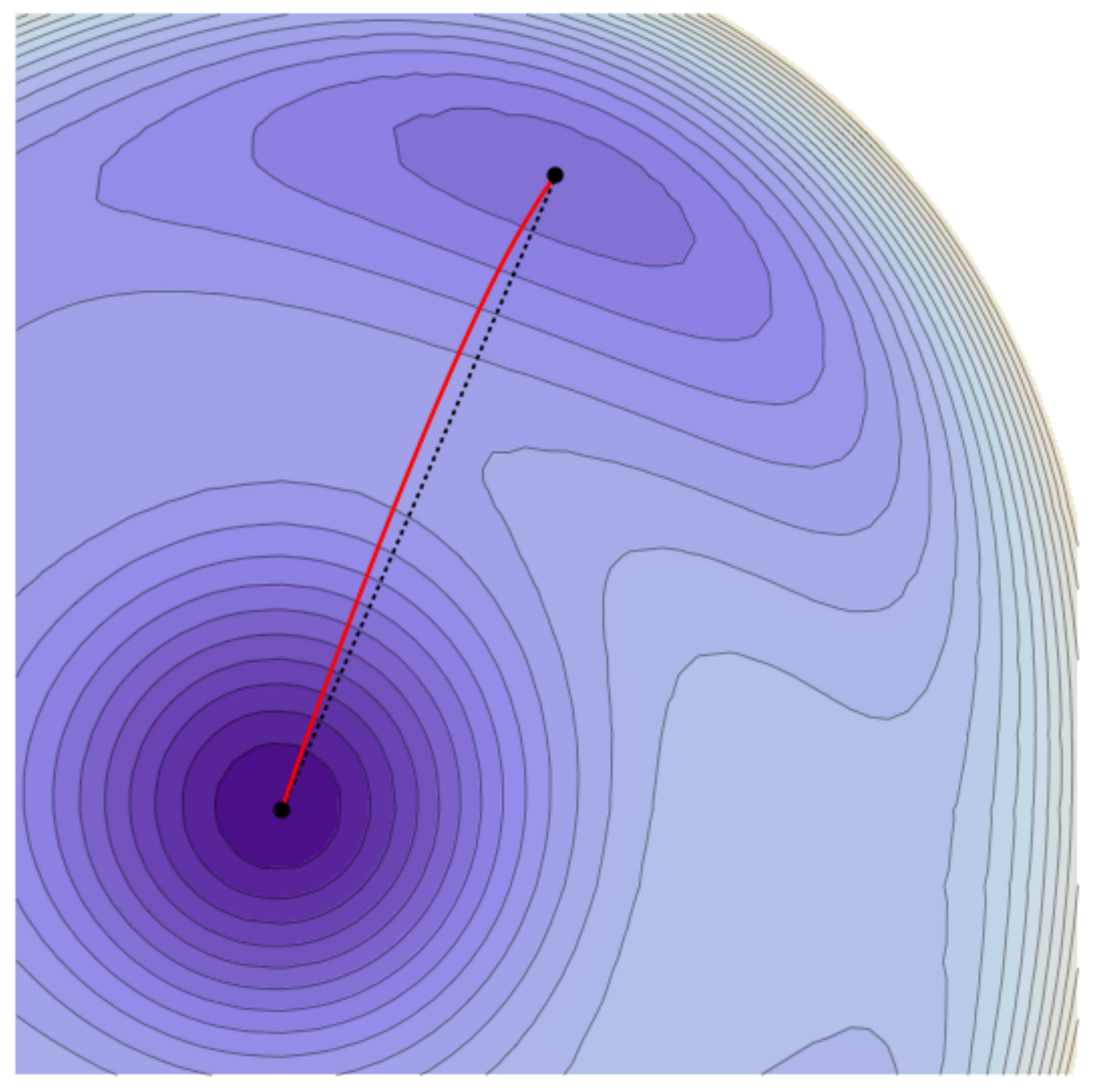}
\caption{\footnotesize{The trajectory (red) in field space taken by the Euclidean instanton solution of \Fig{fig:prof} overlayed on equipotential contours (purple) of the sample $U(1)_B$-violating potential.  To guide the eye, we have drawn a straight dashed line connecting the true and false vacua.  The trajectory taken by the field is not straight; it curves in the $\theta$-direction.}}
\label{fig:Asym}
\end{figure}

\subsection{Asymmetry Generation: Instanton}
\label{subsec:asymmetryNuc}

In the presence of $B$- and $CP$-violating potential terms, the instanton will arc in the $\theta$-direction.  We are interested in computing both the net torque, which fixes the baryon asymmetry in the walls, and the bubble nucleation rate, which sets the percolation time $z_*$.  For both reasons, we need to find the instanton, since it characterizes the most likely bubble configuration to nucleate, and gives the rate via \Eq{decayrate}.   Assuming $SO(4)$ symmetry of the instanton, then the field components $R(\rho)$ and $\theta(\rho)$ are functions of the Euclidean radial variable $\rho$ alone, and the equations of motion for the instanton are
\bea
\label{eq:REOM}
R''+\frac3\rho R' -R\theta'^2&=&\frac{1}{2}\partial_R V \\
\theta''+\frac3\rho \theta'+2 \frac{R'}{R}\theta'&=&\frac{1}{2 R^2}\partial_\theta V.
\label{eq:thetaEOM}
\eea

Boundary conditions are regularity at the origin, so $R'(0)=\theta'(0)=0$, and that far from the bubble the fields settle into their false vacuum values, so $R( \infty ) = R_\text{F}$ and $\theta(\infty)= \theta_\text{F}$, where $\phi_\text{F} = R_\text{F} e^{i\theta_\text{F}}$.   Here, we are assuming that bubble nucleation happens by quantum tunneling through the potential barrier, as would be the case if percolation occurs before reheating.  If instead, bubble nucleation occurs primarily by thermal activation over the potential barrier, then the Euclidean time coordinate is periodic, the $SO(4)$ symmetry becomes $SO(3)\times U(1)$, and the equations of motion change correspondingly \cite{ThermBubble}.

 The field value at the center of the bubble is near, but not exactly at, the true vacuum.  Solutions are found by adjusting the field value at $\rho=0$ so that the boundary conditions are satisfied at $\rho\rightarrow\infty$; that is, we apply Coleman's overshoot/undershoot algorithm generalized to two scalar field directions. A sample instanton, and its curved trajectory through field space, are shown in Figs.~\ref{fig:prof} and \ref{fig:Asym}.

To estimate the extent of the curving, consider a simple potential set by only two parameters:  $m$, the characteristic mass scale of the potential in the $R$-direction, and $\epsilon$, a dimensionless parameter that characterizes the degree of the $B$-violation.  Parametrically,
\bea
\epsilon\sim\frac{\partial_\theta V}{V}.
\eea
Because $m$ is the only dimensional parameter, $R_\text{F}\sim m$, and the instanton solution varies in $\rho$ on scales of order $\rho \sim m^{-1}$.  The parametric scaling of \Eq{eq:thetaEOM} is then such that
\bea
\label{eq:thetascale}
\theta'(\epsilon\sim m^{-1}) &\sim&\epsilon m \\
 \Delta\theta \equiv \theta(\infty)-\theta(0) &\sim&\epsilon.
\eea
To determine the ${\cal O}(1)$ factors here requires finding the instantons numerically as above.

\begin{figure}[t]
\centering
\includegraphics[scale=.97]{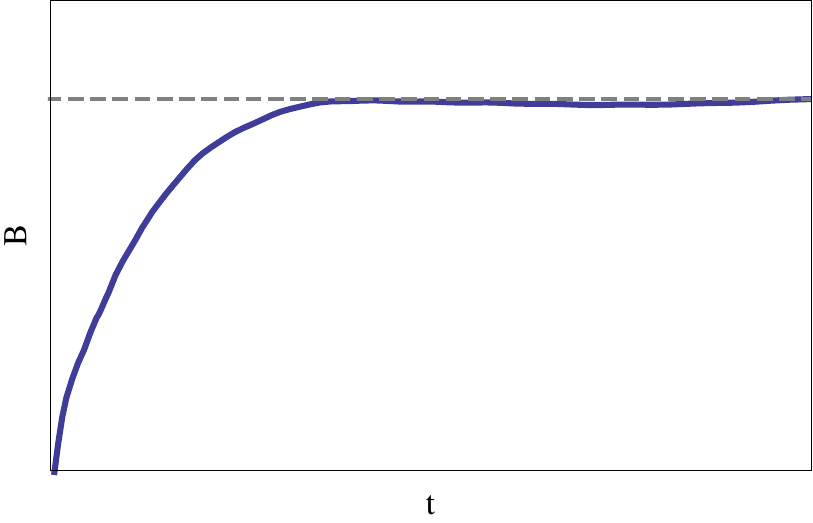}
\caption{\footnotesize{The total $B$ that results from a numeric simulation of an expanding 1+1-dimensional domain wall.  At nucleation, the bubble wall has $B=0$; but as the wall accelerates, $B$ rapidly asymptotes to $\mu^2$, indicated by the dashed gray line.  In higher dimensions than 1+1, total $B$ would grow with the surface area of the bubble, but in 1+1-dimensions, `surface area' is constant.  The simulation was run in an expanding FRW background; gravitational expansion does not affect the value of $\mu^2$.}}
\label{fig:flatmusquared}
\end{figure}

Given instanton profiles $R(\rho)$ and $\theta(\rho)$, the evolution of the bubble post nucleation follows from analytic continuation of the instanton from Euclidean to Minkowski signature.  The typical size of a bubble at the time of nucleation is much smaller than Hubble, so we can ignore the expansion of the Universe and simply continue $\rho\rightarrow\sqrt{r^2-t^2}$, where $r$ is the radial coordinate away from the center of the bubble, and $t$ is time.  At nucleation, $t=0$ and so $\rho=r$, and the field profile that nucleates is a slice through the center of the instanton.

Because the bubble nucleates at rest, $\dot{\theta}=B=0$.  However, as the wall accelerates outwards, spacetime points in the wall traverse an angle in field space $\Delta \theta$ in less and less time, so the baryon density inside the wall grows and grows.  At the same time, as the bubble expands, the thickness of the wall becomes Lorentz contracted, so the baryon density is supported on a smaller and smaller region.  As we will now show, these two effects cancel at late times, and the accelerating bubble wall asymptotes to a constant surface density $\mu^2$ of baryons.

To compute the baryon asymmetry contained in a single bubble wall, we integrate \Eq{eq:B} on a fixed time slice $t=\tau$ long after nucleation
\bea
B_\text{instanton} = \int_{t=\tau} \; R^2  \dot{\theta}\,  d^3x.
\eea
where we plug in the analytically continued classical instanton profile $R(\rho)$ and $\theta(\rho)$; we are working in the semiclassical approximation where loop corrections to this formula are small. Using spherical symmetry, this becomes
\bea
B_\text{instanton} &=& 4\pi \int_{\tau}^\infty \; R\left(\sqrt{r^2-\tau^2}\right)^2  \dot{\theta}\left(\sqrt{r^2-\tau^2}\right)  r^2 dr \nonumber \\
&=&4\pi\int_0^\infty-R(\rho)^2\theta'(\rho)\tau\sqrt{\rho^2+\tau^2}\;d\rho,
\eea
where the prime indicates a derivative with respect to $\rho$.  In the first line, the integral is taken from $\tau$ to $\infty$ because analytic continuation of the instanton only gives the field profile outside of the light-cone; inside the light-cone, the field relaxes towards the $B$-symmetric minimum, producing negligible baryons.  The second line is obtained by changing integration variable from $r$ to $\rho=\sqrt{r^2-\tau^2}$.  Lastly, we add the approximation that $\tau$ is a long time after nucleation, considerably bigger than the size of the bubble at nucleation.  Since $R^2\theta'$ dies off exponentially at large $\rho$, $\tau\gg\rho$ over the region where the integrand has support, and.  
\bea
B_\text{instanton} &\sim& 4\pi \tau^2\times\int_{0}^\infty -R(\rho)^2\theta'(\rho)d\rho,
\eea
which is the bubble surface area at late times ($4\pi\tau^2$), multiplied by the number of baryons per surface area
\bea
\label{eq:mu2def}
\mu^2&\equiv&\int_{0}^\infty  -R(\rho)^2\theta'(\rho)d\rho=\int_\text{F}^\text{T} \; R^2d\theta,
\eea
Because it is a line integral along the instanton field trajectory, $\mu^2$ is constant in time.
In the spirit of \Eq{eq:thetascale}, a parametric estimate is that 
\bea
\mu^2\sim\epsilon R_\text{F}^{\,2}.
\label{eq:mu2est}
\eea
Note that the sign of $\mu^2$ depends on the direction in which $\theta$ arcs, which in turn depends on the imaginary phases in the potential.

\Fig{fig:flatmusquared} shows the total baryon number of a 1+1-dimensional expanding bubble.  The stationary bubble wall carries zero baryon asymmetry, but as it accelerates up to the speed of light the number of baryons per surface area of the bubble wall grows and quickly asymptotes to $\mu^2$.    The integrated number of baryons scales with the surface area; in 1+1 dimensions the `wall' is point-like, and its `surface area' is constant.  

\break

Though we derived \Eq{eq:mu2def} in flat space, it remains true in an FRW background.  Expansion acts globally on the bubble, affecting the growth of proper surface area with time, but it does not act locally---because the bubble wall is much thinner than $H^{-1}$, the field profile is not significantly affected, and so $\mu^2$ remains as in \Eq{eq:mu2def}.  Though the scale factor $a(t)$ does modify the equation of motion, its effect is far smaller than the gradient and potential terms, which set the shape of the wall \cite{VilenkinBasu}.

Because the baryon asymmetry scales with the surface area of the bubble wall rather than the volume, the baryon number density from a single bubble dilutes with time, vanishing as $\tau \rightarrow \infty$.  The baryon asymmetry that is produced is carried away to infinity by the accelerating bubble walls.  Thus, to explain the observed asymmetry, there must be many bubbles, and the bubbles must percolate---not only to ensure that $\phi$ is in the $B$-symmetric minimum today, but also to preserve the asymmetry we have generated from escaping to infinity.

\begin{figure}[t]
\centering
\begin{tabular}{c}
\includegraphics[scale=0.8]{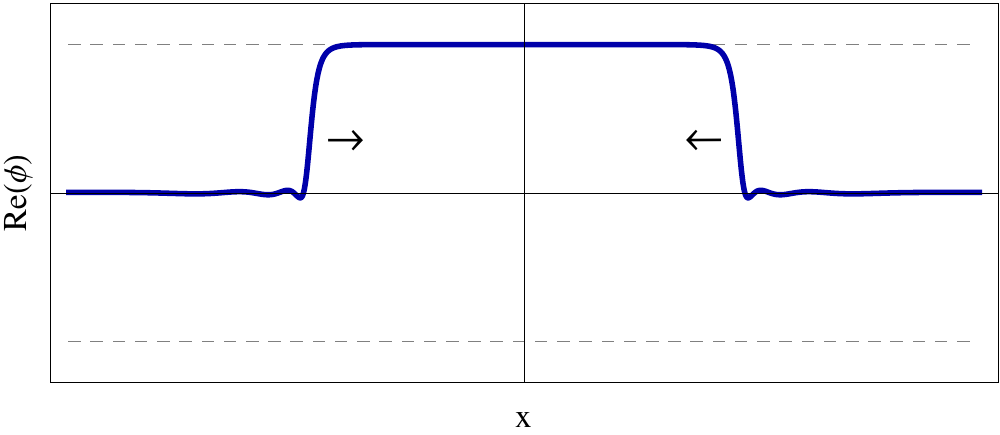}\\
\includegraphics[scale=0.8]{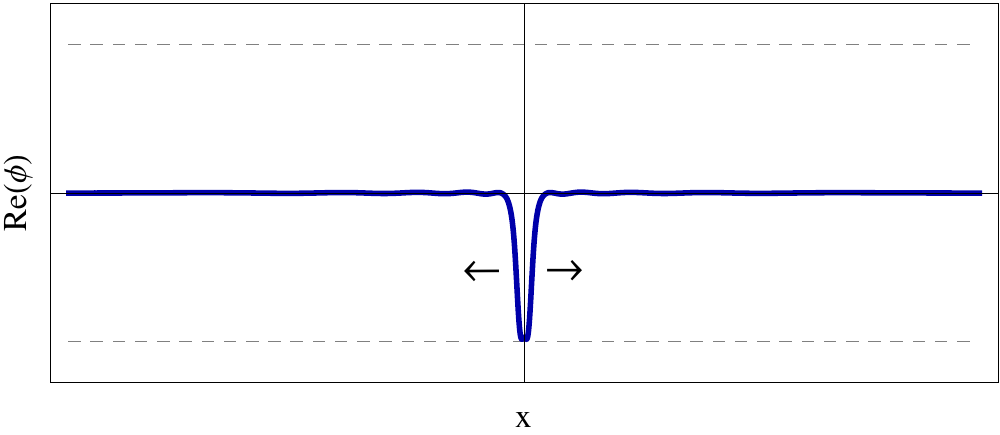}\\
\includegraphics[scale=0.8]{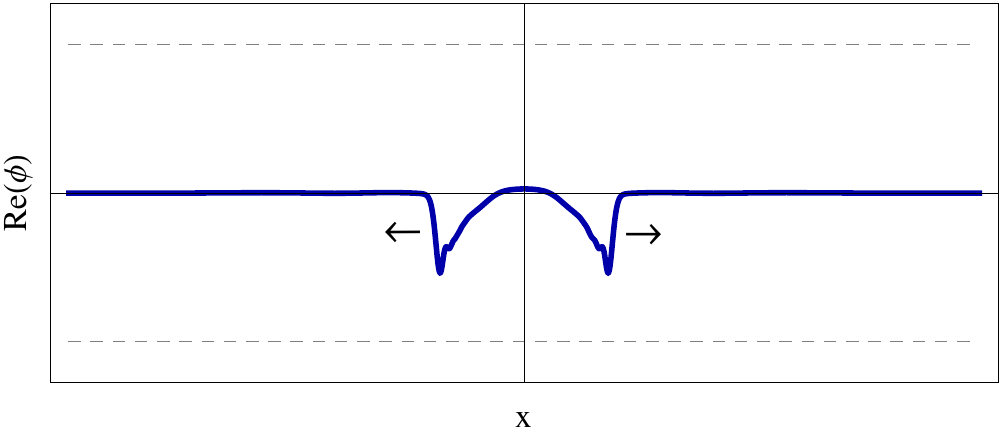}\\
\end{tabular}
\caption{\footnotesize{Sample field profiles of a collision event at three different times. The gray dashed lines indicate the value of $\pm \phi_\text{F}$. Top panel: two boosted domain walls approach one another.  Middle panel: as the walls pass, they solve the \emph{free} wave equation, and so linearly superimpose; the walls pass through each other unimpeded and the field between is deposited at $-\phi_\text{F}$.  Bottom panel: on time scales of order $m^{-1}$, the field begins to respond to the potential.  The field between the walls evolves back towards the true minimum, corresponding to behavior I in the text.}}
\label{fig:Superposition}
\end{figure}

At percolation, there is on average one bubble wall stretched across each Hubble volume.  The expected number of baryons per Hubble volume at percolation, therefore, is the surface area of that wall ($\sim H_*^{-2}$) times the baryon surface density ($\mu^2$).  The number density of baryons right before collisions, therefore, scales like
\bea
n_{B, \text{instanton}}\sim\mu^2\times H_*\sim\epsilon R_\text{F}^{\,2}H_*.
\eea
In the next subsection, we will argue that these baryons are approximately conserved by the collision, so that this $n_{B, \text{instanton}}$ contributes directly to the total $n_B$ in \Eq{eq:totalnB}.

\subsection{Asymmetry Generation: Collisions}
\label{subsec:asymmetryColl}

To gain insight into the dynamics of bubble collisions, we have run numeric simulations for a variety of different models.  Specifically, we have studied the collisions of 1+1-dimensional bubble walls in an expanding, matter-dominated, FRW background with scale factor $a(t)$ for an array of different potentials. 
The equation of motion for $\phi$, ignoring gravitational backreaction, is
\bea
\label{eq:numeric}
\ddot{\phi}+3H\dot{\phi}-\frac{\phi''}{a(t)^2}=-V'(\phi).
\eea
As initial conditions, we used the exact 1+1-dimensional instanton profiles. 

The dynamics of \Eq{eq:numeric} are complicated, but the moment of collision is simple.  The colliding walls are relativistic---they are moving very fast and are very thin, by Lorentz contraction.  This means that the time scale on which they cross is far smaller than both $H^{-1}$ (the time scale on which FRW expansion acts) and $m^{-1}$ (the time scale on which the potential acts).  Gravity and the potential, therefore, can both be ignored, and the field approximately obeys the \emph{free} wave equation  $\ddot{\phi}- \phi''=0$, where we have rescaled $x$ by $a(t)$ the time of collision.  Linear superposition of waves is an exact solution to the free wave equation, so the impinging walls merely pass through one another, and the field in between is deposited at $-\phi_\text{F}$, as shown in \Fig{fig:Superposition}.  This behavior is generic: as long as the walls are moving fast enough, the field value at the intersection of the walls is $-\phi_\text{F}$, independent of the precise shape of the bubble wall or the structure of the potential.

On longer time scales, of order $m^{-1}$, linear superposition is no longer a good approximation and the field begins to evolve under the force of the potential.  The field in between the crossed walls rolls down the potential and begins to oscillate around a local minimum.  There are two behaviors, depending on which local minimum.

\begin{figure}[t]
\centering
\includegraphics[scale=0.285]{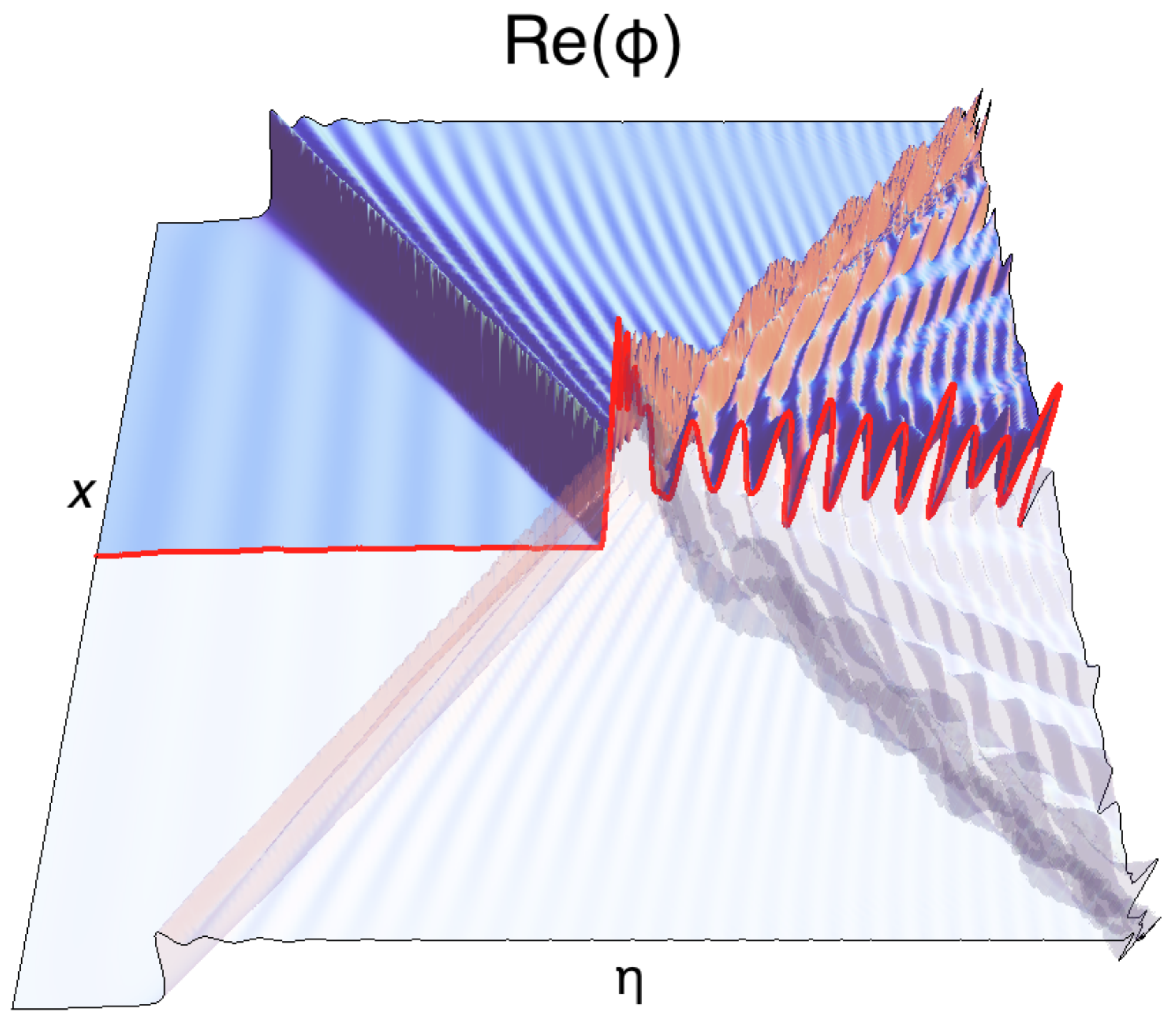} \\
\caption{\footnotesize{$\text{Re}(\phi)$ as function of space and conformal time $\eta$ during a bubble collision showing behavior I.  Using conformal time ensures lightcones travel at 45 degrees.  Domain walls separating the false and true vacua emerge from the top and bottom of the plot and collide at $x=0$.  The walls cross and the field is locally deposited at $-\phi_\text{F}$, from which it evolves back toward the true vacuum.  The value of $\text{Re}(\phi(x=0))$ is shown in red.  Several localized field excitations related to oscillons are visible at late times.}}
\label{fig:RePhiOsc}
\end{figure}

\begin{figure}[t]
\centering
\vspace*{0.07in}
\includegraphics[scale=0.285]{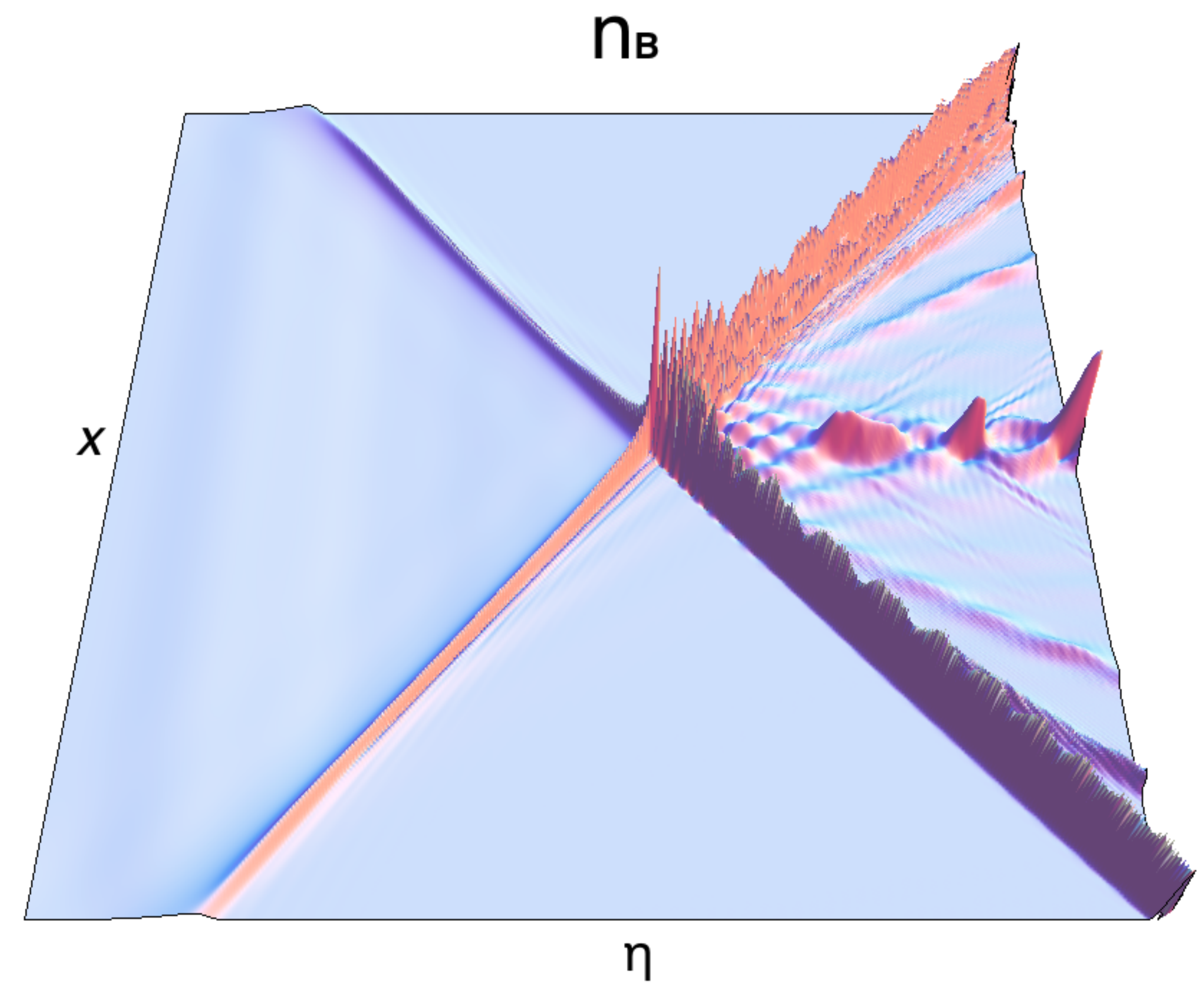} \\
\vspace*{.085in}
\caption{\footnotesize{The baryon number density as a function of space and conformal time $\eta$ for the same collision as \Fig{fig:RePhiOsc}.  Initially, the asymmetry is carried entirely on the walls; as they accelerate the baryon density at the center of the wall grows and the thickness of the wall shrinks. At the collision, the walls move through each other, roughly conserving baryon number, and the evolution of the field from $-\phi_\text{F}$ back to the origin generates new baryons around $x=0$.  The oscillons of \Fig{fig:RePhiOsc} carry non-zero $B$.}}
\label{fig:Current}
\end{figure}

{\it I.  Oscillation about the true minimum.}  The field between the walls, deposited on the other side of the potential at $-\phi_F$, is no longer in a vacuum state; under the force of the potential, it evolves back towards the origin, as shown in the bottom panel of \Fig{fig:Superposition}.  By the time the field reaches the true minimum, it has lost enough kinetic energy to gradients and Hubble friction that it cannot escape; it oscillates, Hubble friction damps the oscillations, and eventually the field settles into the true minimum.  
\Fig{fig:RePhiOsc} shows a collision that illustrates this behavior, and \Fig{fig:Current} depicts the corresponding baryon number density.   

\begin{figure}[t]
\centering
\includegraphics[scale=.91]{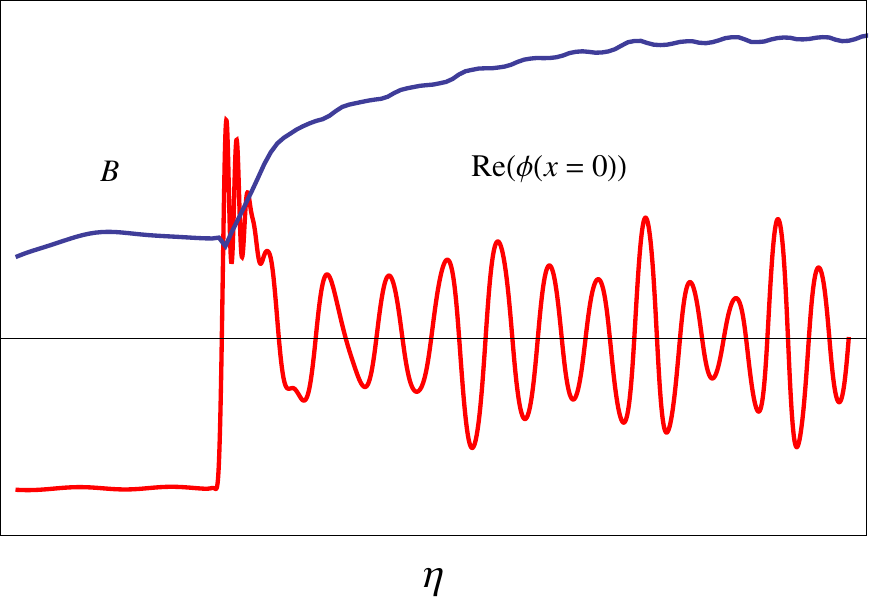} \\
\caption{\footnotesize{The total baryon number $B$ as a function of conformal time for the same collision as  \Fig{fig:RePhiOsc}.  Superimposed, in red, is a plot of $\text{Re}(\phi(x=0))$ during the collision.  $B$ is flat before the collision, as in \Fig{fig:flatmusquared}.  At the collision, the field is deposited at $-\phi_\text{F}$ and evolves back towards the origin; during that evolution, $B$ surges.}}
\label{fig:Bplot}
\end{figure}

As the field at the collision site evolves back towards the origin, it moves through a $B$-violating region of the potential, so a second wave of baryon generation is taking place. \Fig{fig:Bplot} shows the integrated baryon number as a function of time, which in $d+1$-dimensions is
\bea
B(t)=a(t)^d \int  n_B(t,x)\, d^dx.
\eea
Before the collision $B(t)$ was constant, as in \Fig{fig:flatmusquared}, but it makes an abrupt jump upwards at the moment of collision.  The evolution of the field from  $-\phi_\text{F}$ generates new baryons inside the collision lightcone, visible in \Fig{fig:Current}.  A simple estimate can be made for the baryon number generated during this evolution.

The process in which the field in the collision region evolves from $-\phi_\text{F}$ to the origin can be thought of as a localized Affleck-Dine condensate forming and dissolving at the collision site.  If the field takes a time $\Delta t$ to evolve from $-\phi_\text{F}$ to $0$, then the spatial width of this condensate is  of order $\Delta t$, since the bubble walls propagate at nearly the speed of light.  The Affleck-Dine mechanism, were it to occur in this potential, would generate a number density of baryons which scales like $R^2\dot\theta\sim \epsilon R_\text{F}^2   m$.  Multiplying this by the width of the condensate and by the surface area of the collision site per unit volume at percolation $H_*$ gives the expected number density of baryons generated by the collision as
\bea
n_{B, \text{collision}}\sim \epsilon R_\text{F}^2 m  \times \Delta t  H_*\sim\epsilon R_\text{F}^{\,2}H_*,
\eea
where $\Delta t \sim 1/m$ since $m$ is the characteristic scale of the potential.  The contribution to $n_B$ from bubble collisions has the same parametric dependence as the contribution from the instanton.

After the collision, the energy in the field and the walls dissipates, and the asymmetry spreads; the way in which this happens is potential-dependent.  Potentials which grow more slowly than quadratically at the origin, admit two related non-topological solitons that can temporarily trap energy and baryon number \cite{Dymnikova:2000dy, Johnson:2011wt}.  Field excitations that move solely in the $\theta$-direction, locally orbiting the origin, are called $Q$-balls \cite{Coleman:1985ki, Lee:1991ax}; their charge contributes a centrifugal term to their effective potential that makes them absolutely stable.  Field excitation that move solely in the $R$-direction, locally oscillating along a line through the origin, are called oscillons \cite{Segur:1987mg}; these excitations are long-lived, but not eternal.  Our collisions produce a hybrid: it oscillates predominantly in the $R$-direction, and can therefore be seen in \Fig{fig:RePhiOsc}, but it also carries non-zero baryon number, and can therefore be seen in \Fig{fig:Current}.  In other words, the field is locally executing very elliptical orbits about the origin.  A number of these hybrids are visible: a stationary one emerges from the collision site, and several boosted ones fall off the wall as it propagates.  Because these non-topological field configurations probe larger field amplitudes, they are still sensitive to $U(1)_B$ violation, as we discuss in the next subsection.

\break

{\it II.  Pockets of a false vacuum.}  For certain potentials instead of oscillating around the true vacuum, the field at collision site ends up in a false vacuum. This can happen in two ways.  First, if the potential at $-\phi_\text{F}$ is very sloped, then the field can overshoot the true minimum and land back in the original false vacuum, as was noticed in the earliest simulations of bubble collisions \cite{HawkingMoss}.  Second, if the potential happens to have an additional local minimum at $-\phi_\text{F}$, then the field never has to evolve anywhere, since it is already in a false vacuum \cite{Giblin:2010bd}.  The presence of an additional minimum at $-\phi_\text{F}$ does not necessarily require tuning; potentials with approximate $U(1)_B$ have this feature automatically.

\begin{figure}[t]
\centering
\includegraphics[scale=0.285]{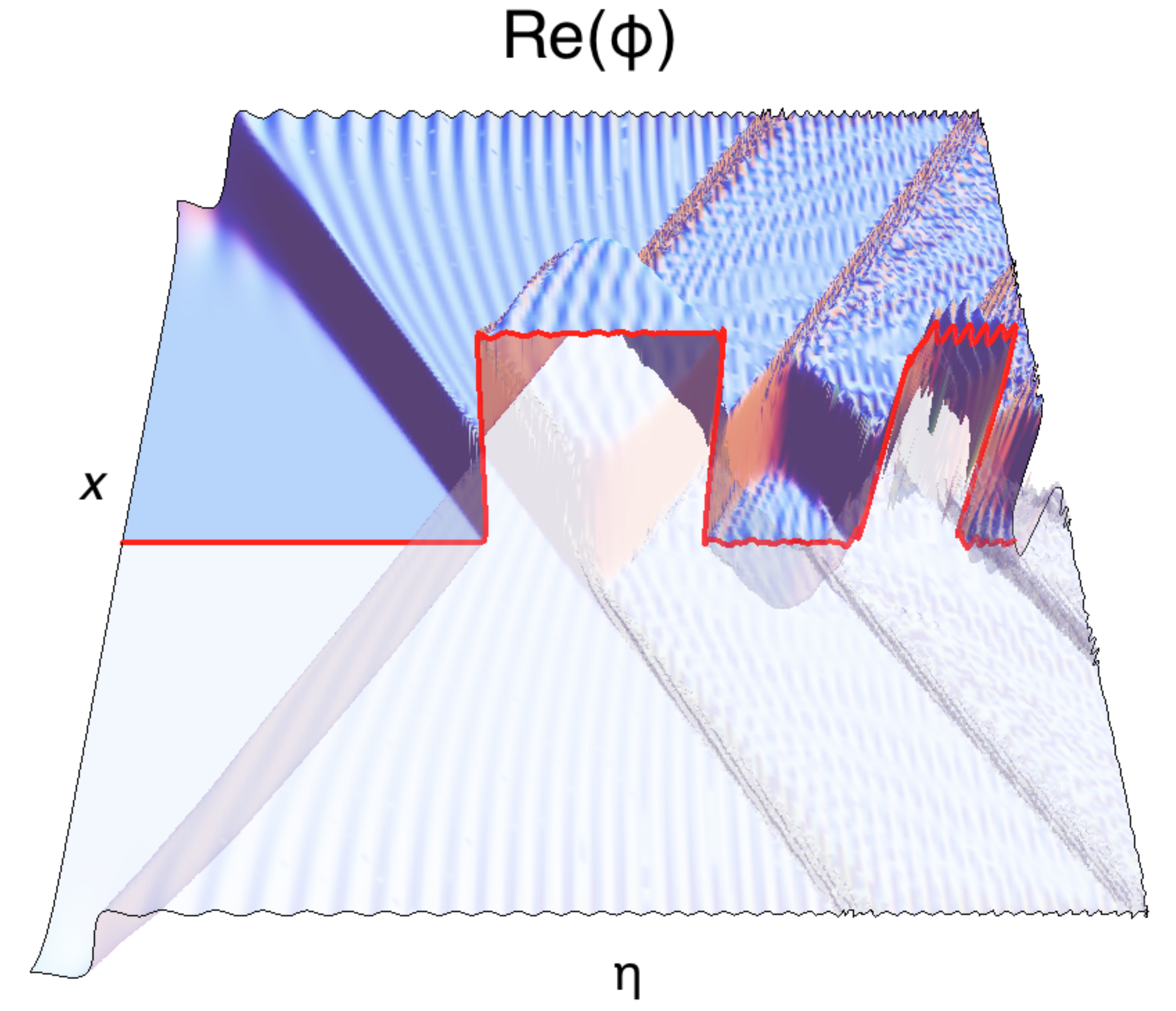} \\
\caption{\footnotesize{$\text{Re}(\phi)$ as function of space and conformal time $\eta$ during a bubble collision showing behavior II. As in \Fig{fig:RePhiOsc}, domain walls separating the false and true vacua emerge from the top and bottom of the plot and collide at $x=0$.  At the collision, the field is deposited at $-\phi_\text{F}$, which is another false vacuum of the theory.  The field oscillates about this vacuum, while the bubble walls move outward, slow, turn around and re-cross.  This process can occur several times.}}
\label{fig:REphipockets}
\end{figure}

The false vacuum provides a locally stable state for the field, around which it can execute small oscillations.  Though locally stable, the field does not remain in the false vacuum forever. 
 The walls, moving away from the collision, now have true vacuum on the outside and false vacuum on the inside.  This induces a pressure that pushes the walls back towards the collision site, so the walls eventually slow, turn around, and re-cross on a time scale of order $H_*^{-1}$, which is far longer than the time scale of oscillations about the minimum.  The formation and collapse of long-lived pockets of false vacuum is shown in \Fig{fig:REphipockets}.  While the field lingers in the false vacuum, $B$ is not conserved, and the fact that the field is oscillating around the false vacuum can yield large fluctuations in the baryon asymmetry.  In this case, the asymmetry is presumably still non-zero, but it is difficult to get an analytic handle on it, and numerical simulations are required.

For potentials with broken $U(1)_B$, there is no reason for $-\phi_\text{F}$ to be a local minimum, and the steepness of the potential at $-\phi_F$ determines which behavior occurs. For potentials with approximate $U(1)_B$, the physics depends on the nature of the false vacuum.  When $-\phi_\text{F}$ has a strong basin of attraction, which happens at larger $z_*$, collisions tend to exhibit behavior II; when $-\phi_\text{F}$ has a weak basin of attraction, which happens at smaller $z_*$, collisions tend to exhibit behavior I.  In the toy model of \Sec{sec:toy}, percolation tends to happen at smaller $z_*$.

\subsection{Washout and Decay}
\label{sec:washout}

After percolation, a baryon asymmetry density of order  $\sim \epsilon R_\text{F}^{\,2}H_*$ is inhomogeneously distributed throughout the Universe in the form of the $\phi$ field.  In order to explain observation, this asymmetry must persist and it must migrate to the standard-model sector.

For the asymmetry to persist, we must avoid two types of washout: classical and thermal.  Classical washout refers to depletion of the asymmetry by evolution under the classical equations of motion from $B$-violating operators present in the early Universe.  The dynamics of classical washout depend on the dimensionality of these operators.  In the case of higher-dimension operators, washout can only be effective at large field values.  But, since the expansion of the Universe and the growth of the bubble both tend to damp field excitations toward the origin, classical washout from higher-dimension interactions is typically evaded, as is true in the Affleck-Dine mechanism.   In the case of marginal or super-renormalizable operators, classical washout may be active even as the fields damp to origin.  Consider, for instance, the case in which the potential for $\phi$ near the origin is of the form $m^2( |\phi|^2 +\epsilon \phi^2 + \textrm{h.c.})$, where $\epsilon$ is a dimensionless measure of $B$ violation.  The $\epsilon$ term induces an `ellipticity' to the potential that splits the mass eigenstates, causing the field to precess as it orbits the origin.  The total baryon asymmetry, therefore, oscillates around its initial value.  The frequency of these oscillations scales like $\sim(\epsilon m)^{-1}$.  Whether the baryon number is spread out evenly, or localized in non-topological solitons, the effect of classical washout is to make it oscillate.

For small $\epsilon$, this precession frequency is far lower than the characteristic oscillation time of the field $m^{-1}$; the field must sit at the origin through many oscillations if the asymmetry is to appreciably change.  As long as the $\phi$ condensate decays before this time, the asymmetry is preserved.  Besides, even if it does not decay in time, the asymmetry oscillates around its initial value, it does not damp; unless there is a conspiracy between the decay time and the oscillation time, the final asymmetry will be an ${\cal O}(1)$ fraction of the initial asymmetry.  Alternatively, in certain models $B$-violation is ${\cal O}(1)$, in which case numerical simulation is necessary to evaluate the degree of washout.

Thermal washout refers to depletion of  $B$ that occurs after the asymmetry is absorbed into the plasma of the early Universe, through scattering processes which involve $B$-violating interactions.  Such scattering tends to restore chemical equilibrium and therefore deplete the asymmetry.  Even if these interactions arise from higher-dimension operators, they can still be significant since the associated scattering rates grow with temperature.  However, as we will discuss in \Sec{sec:toy}, our models evade thermal washout because $B$-violation is sourced by interactions between $\phi$ and the inflaton.  $B$-violation occurs when the temperature is small, before the inflaton has decayed; when the inflaton decays and the temperature becomes large, the $B$-violating interactions have shut off and the asymmetry is frozen in.  Thermal washout is avoided because $B$-violating interactions and the thermal plasma  are never present at the same time.

Finally, for the baryon asymmetry to migrate to the standard-model sector, the $\phi$ condensate must decay.  A direct coupling of $\phi$ to some operator comprised of standard-model fields suffices to transfer the asymmetry; the decay rate of $\phi$ depends on the strength of this coupling.  For homogenous fields, the decay rate of a condensate is more or less the same as the decay rate of $\phi$ quanta in the vacuum \cite{PhiDecay}.  In bubble baryogenesis, however, the field configuration is highly inhomogeneous.  In particular, because the bubble walls are boosted to near the speed of light shortly after nucleation, one might worry that these field fluctuations will be long-lived due to a Lorentz boost factor.  However, as discussed earlier, after the collision these boosted field configurations are not solutions to the equations of motion so they broaden and dissolve.  As the quanta become softer, they can decay. In certain models of bubble baryogenesis, the decay of $\phi$ can be very fast---much faster than the Hubble parameter at the time of percolation.   In such cases, the $\phi$ condensate decays to particles nearly instantaneously after the nucleation event.  Afterwards, there is no classical field, meaning classical washout is straightforwardly evaded.

 \section{Toy Model}
 
 \label{sec:toy}
 
Our discussion in \Sec{general} was framed rather broadly, so in this section we now study a concrete setup.  A working model of bubble baryogenesis must accommodate the following criteria: \begin{itemize}
\item{Percolation.  The Universe must efficiently transit from the false to true vacuum.  }
\item{Asymmetry.  The theory parameters must accommodate the observed baryon asymmetry.}
\item{Perturbativity.  The effective couplings cannot blow up and all energy scales are bounded by the cutoff.}
\item{No Washout.  $B$-violating effects, classical or thermal, must be under control. }
\end{itemize}
In this section we present a toy model which satisfies these criteria.   As we will see, despite its simplicity the toy model may actually be phenomenologically viable.

 \subsection{Model Definition}

Our toy model is defined by a potential of the form
\bea
V(R, \theta) &=& V_0(R) + \epsilon V_1(R,\theta),
\eea
where $\epsilon$ is a small parameter characterizing $B$- and $CP$- violating interactions.  The $B$-symmetric part of the potential is
\bea
V_0 &=& m^2|\phi|^2 - A |\phi|^3 + \lambda |\phi|^4.
\label{eq:Vsym}
\eea
In general, the shape of the potential will vary in time due to couplings between $\phi$ and the inflaton $\chi$, whose vacuum expectation value is time-dependent.
This occurs in Affleck-Dine models of baryogenesis, where such couplings induce Hubble-dependent parameters in the action.  For simplicity, we take $\lambda$ and $A$ to be  constant, but
 \bea
 m^2 &=& \tilde m^2 - \frac{\rho}{\Lambda^2},
 \label{eq:Hparams}
 \eea
 where $\rho = 3 H^2 \mPl^2$ is the energy density of the Universe.  Here we require $\Lambda^2 >0$ and $\tilde m^2 > 0$  so that $B$ is spontaneously broken at early times but restored in the present day.
In a supersymmetric context,  such a $\rho$ dependence would originate from 
$|\chi|^2 |\phi|^2/\Lambda^2$ in  the Kahler potential.

If the couplings between $\phi$ and $\chi$ are $B$- and $CP$-violating, then $V_1$ will also contain time-dependent terms.   At least two $B$-breaking operators are required---otherwise all $CP$ phases can be removed by a field redefinition.  As discussed in \Sec{sec:washout}, if $B$ violation is higher-dimensional, e.g.~$\phi^5$ and $\phi^6$, then explicit breaking is localized far from the origin and classical washout is ameliorated by Hubble damping of the field.  In contrast, renormalizable $B$-violating operators, e.g.~$\phi^2$ and $\phi^3$, typically mediate classical washout with model-dependent effects.
Importantly, we assume that $B$-violating interactions are sourced by the inflaton alone, so $B$ is restored after reheating---thus, thermal washout is evaded.  This setup can be easily engineered by invoking additional symmetries under which both $\phi$ and $\chi$ transform.

Let us  outline the cosmological history of this model.  We begin during inflation, when $H$ and therefore all model parameters are constant in time.  During this epoch, $m^2 < 0$ and the field resides at $\phi\neq 0$.  After inflation ends, the inflaton starts to oscillate around its minimum and the Universe shifts from vacuum-energy domination to matter domination.  As $H$ decreases, $m^2$ eventually becomes positive, growing monotonically until the $\phi\neq 0 $ vacuum  becomes metastable.  Once the nucleation rate rises sufficiently, percolation occurs.   The $B$- and $CP$-violating interactions in the potential cause the nucleated bubble walls to accumulate baryons; subsequent bubble collisions yield an additional asymmetry component.  

In the subsequent sections, we will analyze the vacuum structure, nucleation rate, and asymmetry generation in this model.

\begin{figure}[t]
\centering
\includegraphics[scale=0.8]{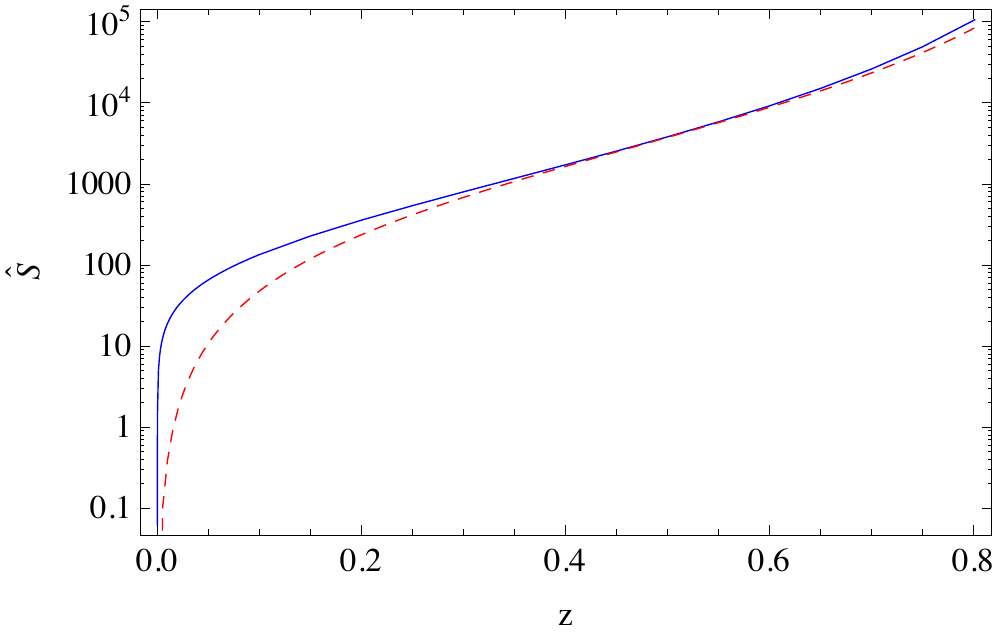} \\
\caption{\footnotesize{$\Delta \hat{S}(z)$ for the potential in \Eq{eq:Vsym}, computed numerically (blue solid) and analytically in the thin-wall approximation (red dashed).}}
\label{fig:SHat}
\end{figure}

\subsection{The Instanton}

\label{sec:toyvac}

Neglecting effects proportional to $\epsilon$, the Euclidean action for this theory is
\bea
S &=& 2 \pi^2 \int \rho^3 d\rho  \left[  \left(\frac{\partial R}{ \partial \rho}\right)^2 + m^2 R^2  - A R^3 + \lambda^2 R^4  \right]. \nonumber \\
\eea
While the model parameters vary in time, they do so on scales $H \ll 1/m$, so we treat them as constant in our analysis of the instanton.
It is convenient to transform to dimensionless variables,
\bea
S&=& \hat S /\lambda^2  \nonumber \\
R &=& \hat R  m / \lambda \nonumber \\
A &=& \hat A  m\lambda \nonumber \\
\rho &=& \hat \rho / m,
\label{varchange}
\eea
where 
\bea
\hat{S} &=& 2 \pi^2 \int \hat \rho^3 d \hat \rho \;  \left[ \left(\frac{ \partial \hat R}{ \partial \hat \rho}\right)^2 +\hat R^2 -\hat A \hat R^3 + \hat R^4 \right].
\label{Snewvars}
\eea
As a result of our change of variables, the rescaled Euclidean action $\hat S$ is a function of $\hat A$ alone.

As discussed in \Sec{subsec:vacuum}, the variable $z$ is a convenient reparametrization of time in which $1>z>0$ corresponds to the epoch in which tunneling is allowed.  We can express $\hat A$ in terms of $z$ as
\bea
\hat A \left(z\right)&=& \sqrt{32/9} + z(2-\sqrt{32/9}),
\label{eq:Ahat}
\eea
where $1>z>0$ maps onto the range $2 > \hat A > \sqrt{32/9}$.
 Within this interval, there is a true and false vacuum located at 
  \bea
\hat R_\text{T} =0 &, & \hat R_\text{F} =  \frac{3\hat A}{8} \left(1- \sqrt{1-\frac{32}{9\hat A^2}}\right).
\label{eq:rF}
\eea

Next, we evaluate $\Delta \hat S$ by solving the associated Euclidean equation of motion for an $SO(4)$ symmetric ansatz $\hat R$ subject to the initial condition $\partial \hat R / \partial \hat \rho=0$ at $\hat \rho =0$. Solving for $\Delta \hat S$ numerically, we find that for $z\lesssim 0.25$, $\Delta \hat S$ is very well fitted by the function
\bea
\Delta \hat S &=& 431.5 z^{0.679}+8139.4 z^{2.27}
\label{numS}
\eea
As $z\rightarrow 0$ the phase transition shifts from first-order to second-order.  As $z \rightarrow 1$ the bubble becomes thin-walled.  From \Fig{fig:SHat}, it is clear that our numerics agree with analytic expressions in this regime.

The determinant prefactor can also be straightforwardly estimated.  From \cite{ColemanQM} we can compute the $K$ factor in \Eq{decayrate}
\bea
K &=& \frac{\Delta S^2}{4\pi^2} \left| \frac{\det' (-\Box + V''(\phi(\rho)))}{\det(-\Box + V''(\phi_\text{F}))} \right|^{-1/2} \\
&\sim&  \frac{\Delta \hat S^2 m^4}{4\pi^2\lambda^4},
\label{det}
\eea
where $\phi(\rho)$ is the instanton solution and $\det'$ indicates the determinant with zero modes removed. In the second line we used that $\rho$ varies on scales of order $m^{-1}$ to estimate the determinant factors.

\begin{figure}[t]
\centering
\includegraphics[scale=0.65]{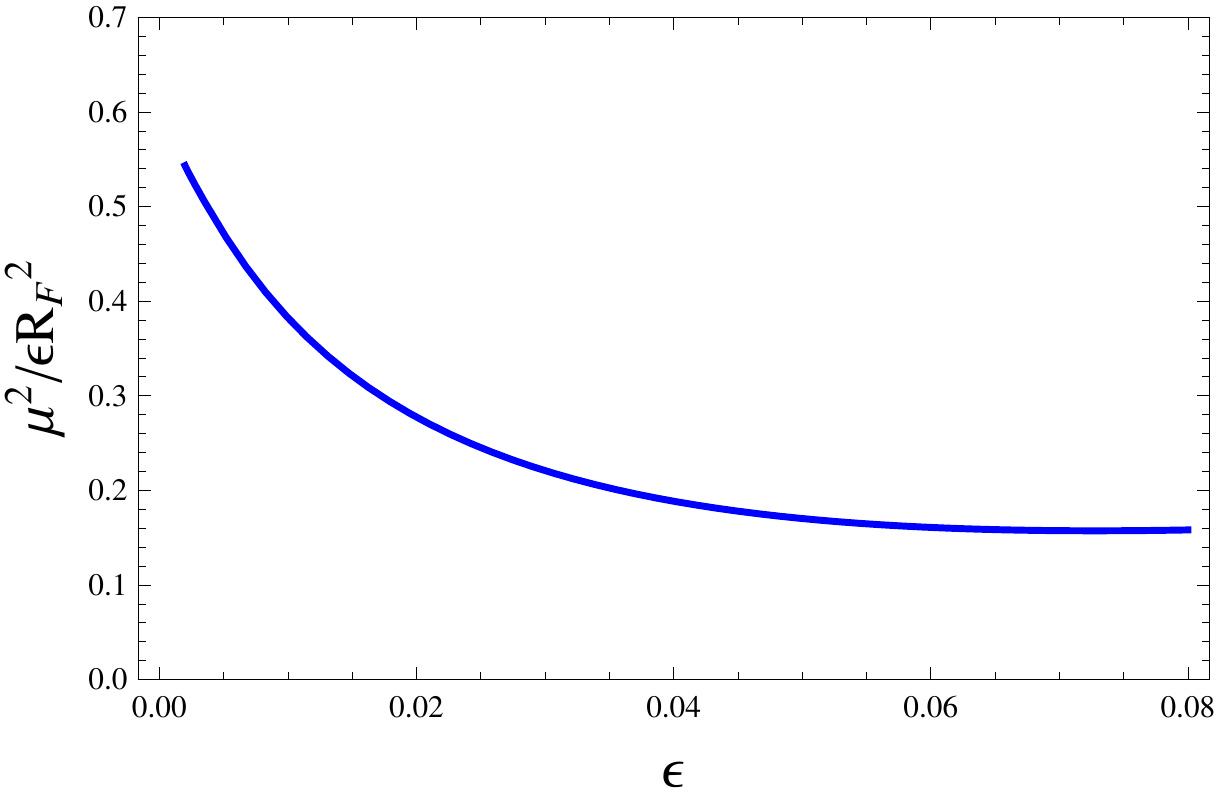} \\
\caption{\footnotesize{Plot of the exact value for $\mu^2$ divided by
our parametric estimate, $\epsilon R_\text{F}^2$---the answers agree
within ${\cal O}(1)$ factors.  Here we used $V_0$ corresponding to $z=0.2$ and
$V_1=R^2\cos(2\theta+\pi/4)+R^3\cos(3\theta)$.}}
\label{fig:muEpsilon}
\end{figure}

Up till now we have neglected $B$- and $CP$-violating effects proportional to $\epsilon$.  Employing our numerical code, we have generated instanton profiles for the potential at finite $\epsilon$. Given these numerical solutions, we can compute $\mu^2$, defined in \Eq{eq:mu2def} as the measure of the baryon asymmetry in the bubble wall.  As shown in \Fig{fig:muEpsilon}, at small $\epsilon$, these numerical results match the simple estimate for $\mu^2$ described in \Eq{eq:mu2est}.

\subsection{Before Percolation}

Consider the cosmological history of this model leading up to percolation.
As shown in \Sec{sec:toyvac}, the vacuum structure of the theory depends solely on $\hat A$, which varies in time with the energy density, $\rho$.     
Shortly after inflation, $m^2 < 0$ and the potential has a single minimum at large field values.
As the universe cools, eventually $m^2>0$ and an additional local minimum forms at the origin---at this point $\hat A$ is divergent.  As $m^2$ continues to increase, $\hat A$ monotonically decreases, and a first-order phase transition becomes possible in the window $2 > \hat A > \sqrt{32/9}$.

Plugging $\rho = 3 H^2 \mPl^2$ into \Eq{eq:Ahat} and \Eq{eq:Hparams}, we obtain expressions for the important physical quantities as a function of $z$,
\bea
m(z) &=& \frac{A}{\lambda \hat A(z)}\\
H(z) &=& \frac{ \Lambda \sqrt{\tilde m^2 - m(z)^2}}{\sqrt{3}\mPl},
\label{zfunctions}
\eea
where $\hat A(z)$ is defined in \Eq{eq:Ahat}.  Given the parametric dependences in \Eq{zfunctions}, it is clear that during the first-order phase transition, the dimensionful parameters, $H$, $m$, and $A$ are all well below the cutoff $\Lambda$ and so the effective theory description remains valid.

Eventually, the universe cools sufficiently that Hubble decreases enough that there is a first-order phase transition. Percolation occurs when \Eq{percApprox} is satisfied, which is roughly when $\Gamma_* \simeq H_*^4$; in the context of our model this approximation is accurate to $\lesssim 15\%$.  Recall that $\Gamma_*$ implicitly depends on $z_*$ through the model parameters in \Eq{zfunctions}.  The criterion for percolation can be rewritten as
\bea
H_*^4 &=& \frac{\Delta S_*^2 m_*^4}{4\pi^2} e^{-\Delta S_*}.
\eea
The exponential factor essentially fixes the solution to this equation, and so the prefactors are only logarithmically important.  Solving for $S_*$ yields
\bea
\Delta S_* &= & \log \left(\frac{9 \mPl^4 \Delta S_*^2}{4 \pi^2 \Lambda^4 (1-\tilde m^2 / m_*^2)^2}\right) \\
&\sim & 4 \log \left(\frac{\mPl}{ \Lambda}\right),
\eea
where the ellipses denote logarithms of ${\cal O}(1)$ numbers.   If $S_* \ll 1$, then the nucleation rate is very  high at the onset of percolation, indicating that the phase transition is bordering on second order.  While this is not necessarily bad in and of itself, in this regime $\phi$ evolves like a slow roll Affleck-Dine condensate.  Of course, we will be interested in the first-order regime, whereby tunneling is the dominant mode of the phase transition.  To achieve this, we require a modestly sized instanton action, so $\Lambda \ll \mPl$, which is to say that the higher dimension operators coupling the inflaton to $\phi$ cannot be Planck slop operators, and must be suppressed by a lower scale.

In this toy model $z_\star$ is typically small and we are in the weakly first-order regime. Allowing both $A$ and $\lambda$ to also vary with time can yield any value of $z_\star$ between $0$ and $1$.

\subsection{After Percolation}

Once percolation occurs, bubbles of true vacua nucleate and soon fill the volume of space.  Using \Eq{eq:mu2est}, which applies for $\epsilon \ll 1$, we find that the surface density of baryons on the walls is
\bea
\mu_*^2 \sim \frac{\epsilon  m_*^2  \hat R_{\text{F}*}^2   }{\lambda^2}.
\eea
The total baryon number density $n_B$ arising from the initial instanton plus the subsequent bubble collisions is given in \Eq{eq:totalnB}.
To compute the observed baryon asymmetry today, we need to consider the remainder of the cosmological history.  Because the inflaton sources the $B$- and $CP$- violating interactions of $\phi$, we require that the decay of the inflaton, and thus reheating, occur after the percolating phase transition.  Consequently, the asymmetric yield at the time of reheating is given by
\bea
\eta_B &\equiv & \frac{n_{B}}{s_{R}}\frac{H_\text{R}^2 }{H_*^2}\\
&\sim& \frac{\sqrt{3}\epsilon  m_*  T_\text{R} \hat R_{\text{F}*}^2}{4\lambda^2 \mPl \Lambda \sqrt{\tilde m^2/m_*^2-1}},
\label{eta}
\eea
where  $H_\text{R}$ and $T_\text{R}$ are the Hubble parameter and the temperature, respectively, at the time of reheating.   
The above estimates neglect the effects of classical washout, but as  discussed in \Sec{sec:washout} these effects are expected to change the asymmetry by an ${\cal O}(1)$ fraction and are model-dependent. 

\begin{figure}[t]
\centering
\includegraphics[scale=0.875]{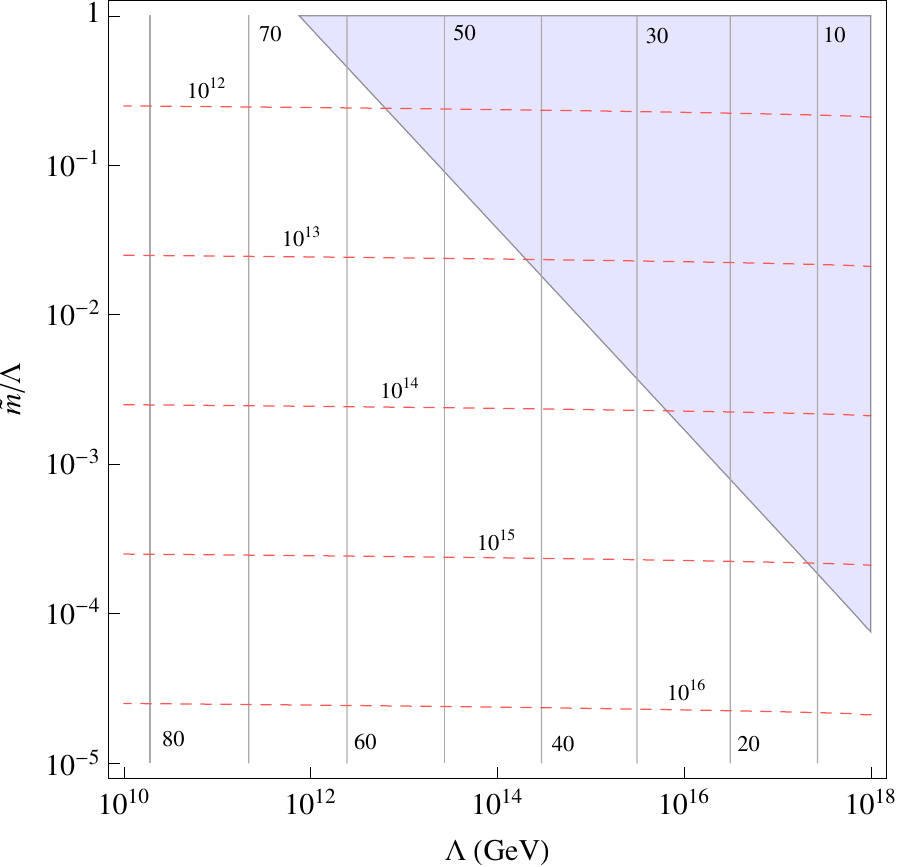} \\
\caption{\footnotesize{Parameter space for the toy model, fixing $\eta_B$ to the observed baryon asymmetry today.
The blue region depicts the allowed region after requiring that reheating occurs after percolation, $\rho_\text{R} \lesssim \rho_*$. The red dotted lines are contours of $T_\text{R}$ and the solid gray lines are contours of $\Delta S_*$. We have fixed $\lambda =1$, $\epsilon = 0.1$, $\tilde m = A$. }}
\label{fig:ParamSpace}
\end{figure}

To accommodate the present day observed baryon asymmetry, we require that $\eta_B \sim 6 \times 10^{-10}$.  The energy density of the Universe at reheating is bounded from above by the energy density at percolation, so $T_\text{R}^4 \sim \rho_\text{R} \lesssim \rho_*$.  In turn, this places an absolute upper limit on $\eta_B$ from \Eq{eta}.
Given the observed baryon asymmetry, this upper bound can be rephrased as a lower bound on $\tilde m$ in terms of the other fundamental parameters
\bea
\tilde{m} &\gtrsim& \Lambda\left(\frac{10^{10} \textrm{ GeV} \times \lambda^2}{\epsilon\Lambda}\right)^{2/3},
\label{eq:massbound}
\eea
where for simplicity we have dropped the ${\cal O}(1)$ factors from the $z_*$ dependence.
This limit is a substantial constraint on $\tilde m$ in this toy theory, and is depicted in the blue region of \Fig{fig:ParamSpace}.

Next, we briefly discuss how the asymmetric yield in \Eq{eta} is actually transferred from $\phi$ into standard model fields.  In a supersymmetric context this is achieved, for example, by the operator $UDD/M$ in the superpotential,
where $M$ is the mass scale of some connector field which has been integrated out.   This is the lowest-dimension operator which can link standard model $B$ to a gauge singlet field $\phi$.   Given this operator, the field $\phi$ has a decay rate of 
\bea
\Gamma(\phi \rightarrow qqq) &\sim& \frac{1}{128  \pi^3} \frac{m^3}{M^2},
\eea
which can be much greater than $H_*$, the Hubble parameter at the time of percolation.  In this case the $\phi$ field within each bubble of nucleated true vacuum decays very shortly after percolation.  Classical washout from $B$-violating interactions are thus minimized since $\phi$ decays so fast into particle quanta.  Alternatively, fast decays can  occur if $\phi$ decays promptly to additional non-gauge-singlet particles which in turn decay to standard model fields.
Finally, while our discussion thus far has been framed within the context of $B$, the asymmetry can of course be converted into $L$ via the operators $\phi L H_u$, $\phi L L E$, or $\phi Q L D$, or into a dark matter asymmetry via similar interactions.

\section{More Realistic Models}
\label{sec:realistic}

In this section we discuss possible realizations of bubble baryogenesis in more realistic contexts.  The examples here are supersymmetric, but we emphasize that this is not a requirement for bubble baryogenesis.

\subsection{Neutrino Seesaw}

Consider the MSSM augmented by a supersymmetric neutrino seesaw.   The superpotential is
\bea
W &=& \lambda L  H_u N+ \frac{1}{2} M N^2,
\eea
where $N$ denotes the sterile neutrinos and we have suppressed all flavor indices.
Explicit violation of $U(1)_L$ is present in the form of the mass parameter, $M$.  
Integrating out the heavy $N$ fields yields the active neutrino masses, whose sum is bounded by cosmological measurements to be less than $0.17$ eV \cite{Seljak}.  Requiring that $\lambda \lesssim 1$ implies that $M \lesssim 10^{14}$ GeV.

The supersymmetric $F$-term and $D$-term contributions to the potential are
\bea
V_F &=& |\lambda H_u N|^2 +   |\lambda L N|^2 +|\lambda L H_u + M N|^2 \nonumber\\
V_D &=& \frac{g'^2 + g^2}{8} (|H_u|^2 - |L|^2)^2.
\label{eq:FDN}
\eea
If supersymmetry breaking approximately preserves $L$, then the corresponding contributions to the potential are of the form
\bea
\tilde V &=& \tilde V_0 +\epsilon \tilde V_1
\label{softsplit}
\eea
where $\epsilon$ is small and
\bea
\tilde V_{0} &=& m_{L}^2 |L|^2+ m_{N}^2 |N|^2 +m_{H_u }^2 |H_u|^2 \nonumber \\
&&+ A L H_u N  + \text{h.c.}
\label{eq:softN}
\eea
Because the parameters in $\tilde V$  acquire contributions from both zero-temperature supersymmetry breaking and Hubble-induced supersymmetry breaking, they are in general time-dependent.  Which specific contributions arise depends on the symmetry structure of the ultraviolet theory, which dictates the coupling between the MSSM fields, the inflaton, and the supersymmetry breaking sector.  For instance, if the inflaton carries $R$-parity, then $L$-violating terms like $LH_u$ or $N^3$ could be present in $\tilde V_1$.

The scalar potential is complicated, and contains a large number of fields and parameters.   However, it is clear that all the required features of bubble baryogenesis are present.  First, the $A$ parameter may be large in the early Universe and produce global minima far from the origin of field space.  Second, these $B$-breaking vacua are stabilized by the quartic $\lambda$ which can be ${\cal O}(1)$ in order to achieve a sufficiently large nucleation rate.  Third, $L$- and $CP$-violating interactions can torque the Euclidean instanton solution during a first-order phase transition.  Hence, we do not expect bubble baryogenesis in this potential to differ in any qualitative way from the toy model in \Sec{sec:toy}.

A complete analysis of the multi-field potential of the supersymmetric neutrino seesaw would be non-trivial.  However, we take note of certain simplifications to the theory which can reduce the potential to a solvable one.  In particular, the lepton flavor indices serve largely to complicate the analysis, so the potential is greatly simplified by taking a single flavor of $L$ and $N$ to be the only fields active in the dynamics.  Furthermore, the $D$-term contribution in \Eq{eq:FDN} tends to fix $|L| = |H_u|$\footnote{Some cosmological implications of such $D$-flat directions are discussed in \cite{Ainflation}.}, which effectively eliminates another set  of field directions.

\subsection{Color-Breaking Minima}

Bubble baryogenesis may also be possible within the context of tunneling from color-breaking minima \cite{CCB} to the electroweak vacuum within the MSSM.  If supersymmetry-breaking $A$-terms are too large, then deep minima can form at field values away from the origin, inducing an instability for the electroweak vacuum.  The vacuum dynamics are dominated by the field directions of the top squark and the Higgses, where the superpotential is dominated by the usual top quark Yukawa coupling,
\bea
W&=& \lambda_3 Q_3 H_u U_3 .
\eea
Given approximately $B$-symmetric supersymmetry breaking, the corresponding terms in the potential are as in \Eq{softsplit} but
with
\bea
\tilde V_0 &=& m_{Q_3}^2 |Q_3|^2+ m_{U_3}^2 |U_3|^2 +m_{H_u}^2 |H_u|^2  \nonumber\\
&&+ A_3 Q_3 H_u  U_3  +\textrm{h.c.}
\eea
Considering the $D$-flat field direction $|Q_3| =|U_3| = |H_u|$, then the absence of color-breaking minima implies that
\bea
A_3^2 \leq 3 \lambda_3^2 (m_{Q_3}^2 + m_{U_3}^2 + m_{H_u}^2),
\label{eq:Abound}
\eea 
if the electroweak vacuum is to be absolutely stable.  

For our purposes we take the opposite tack---we want the electroweak vacuum to be unstable in the early Universe.   Couplings to the inflaton can be easily arranged so that \Eq{eq:Abound} fails in the early Universe, so that the fields reside in the true, color-breaking vacuum.  As the Universe cools, \Eq{eq:Abound} is eventually  satisfied, and the fields can tunnel from the color-breaking minimum to the electroweak vacuum.  

Of course, to generate a baryon asymmetry, this phase transition requires $B$- and $CP$-violating interactions.  The natural candidate for this is the Hubble-induced cubic term, $U_i D_j D_k$,
whose coupling can carry non-zero $CP$ phases.  Because this operator involves multiple squark flavors, we are then required to understand the tunneling from the color-breaking minimum in other squark directions besides the stop.    We leave a proper analysis of this scenario for future work.

\section{Observational Consequences}
\label{sec:pheno}

The byproduct of bubble baryogenesis is a frothy mixture of standard model baryons, inhomogeneously distributed in the early Universe.  This inhomogeneity provides a potentially dangerous relic, since the observed baryon density is known to be homogeneous at the epoch of big bang nucleosynthesis (BBN).   \emph{On average}, inhomogeneities are small at BBN: the percolating bubbles have a length scale set by $H_*^{-1}$, which is far smaller than $(a_*/a_\text{BBN})H_\text{BBN}^{-1}$, the size of an inhomogeneity at percolation that would grow to be Hubble size at BBN.  The potentially dangerous relic doesn't come from the average bubble, it comes from the rare bubble that nucleates early enough and avoids enough collisions to grow big by percolation.  Constraints from big bubbles were studied in \cite{TWW} in the context of extended inflation \cite{ExtendedInf}, which also features a first-order transition around the end of inflation.  Though the context was different, the constraints are purely geometrical, so the analysis in \cite{TWW} carries over.  Big bubbles constrain the decay rate considerably before percolation to be small, so that big bubbles are exceedingly unlikely.  Bubble baryogenesis is aided in avoiding this constraint by the fact that the nucleation rate can be completely shut off at early times.

Constraints aside, the first-order phase transition also opens up new observational signatures, like gravitational waves.  Bubble collisions are an efficient producer of gravitational waves; numeric estimates in \cite{KTW} show that as much as .1\% of the energy released in the transition can end up in gravitational waves.  Unfortunately, the energy for us is small---most of the energy density of the Universe is in the inflaton---but the gravitational waves have a distinct signature which may make observation feasible.  Because the colliding bubbles at percolation have roughly the same size, the gravitational wave spectrum has a spike at $H_*$ \cite{TW}.  This observational signature is distinct from Affleck-Dine.

Additionally, black holes might form at the collision sites.  This is an intriguing possibility, because a current tension in the data of the mass distribution of early quasars would be alleviated by a source of primordial `seed' black holes \cite{Volonteri:2002vz}.  Also, because these black holes form with a characteristic size, there could be a second bump in the gravitational wave spectrum from their coincident evaporation. 

Lastly, as we discussed in \Sec{subsec:asymmetryColl}, bubble collisions can spawn non-topological solitons, like oscillons and $Q$-balls. Oscillons, though long-lived, are not typically stable on present Hubble time-scales \cite{Farhi:2007wj,Amin:2010jq,Amin:2011hj}; they radiate energy and dissociate.  $Q$-balls, on the other hand, can be stable and persist since they are charge-stabilized \cite{Cohen:1986ct,Q1}. In Affleck-Dine baryogenesis, $Q$-balls typically form in gauge-mediated theories \cite{Q2}, and may or may not form in gravity-mediated theories, depending on the supersymmetric spectrum \cite{Qgrav,Q3}.   Analogous statements are likely true for bubble baryogenesis.  If $Q$-balls form, their subsequent evolution is model-dependent: $Q$-balls that carry $B$ will be absolutely stable if their mass-to-baryon-charge ratio is smaller than that of a proton; $Q$-balls that carry $L$, however, typically decay, since the mass-to-lepton-charge ratio of the neutrino is so small.

\section{Future Directions}
\label{sec:conclusions}

Bubble baryogenesis is a novel scheme for the generation of the cosmological baryon asymmetry. A scalar baryon undergoes a first-order phase transition in the early Universe, and a baryon asymmetry is generated by the process of bubble nucleation and the subsequent bubble collisions.  We have presented an explicit toy model to illustrate the basic features of the mechanism, and introduced a handful of realistic models.  In addition to fleshing out these more realistic theories, there exists a variety of interesting directions for future work.

First, in \Sec{subsec:asymmetryNuc}, we argued that a single bubble could never explain the observed baryon asymmetry because the baryons are carried away to infinity.  This is not strictly true: a loophole is provided in theories with large extra dimensions---the same loophole that boom-and-bust inflation \cite{ABrown} exploits to provide a graceful exit to old inflation.  If a bubble nucleates smaller than the size of the extra dimensions, then as it grows, it wraps the extra dimension and collides with itself on the other side.  The bubble wall no longer runs off to infinity; the self-collision preserves the baryons in the wall, and distributes them uniformly throughout the interior of the bubble.

Second, as we discussed in \Sec{subsec:vacuum} there is the possibility of bubble baryogenesis after reheating.  Thermal effects from the big-bang plasma can induce time-dependent couplings which give rise to a first-order phase transition; the thermal plasma also assists the transition by enhancing the nucleation rate.  To achieve such a scenario requires engineering the appropriate thermal potential and avoiding thermal washout.

Third, models considered in this paper had only a single instanton mediating decay, which imposed a relationship between the efficiency of the phase transition and the resulting asymmetry.  This need not be the case.  Consider a true and false vacuum connected by two instantons: a dominant one that is purely radial so that it generates zero baryons, and a subdominant one that arcs so that it alone is responsible for generating baryons.  Physically, this would correspond to a percolating phase transition in which the vast majority of nucleated bubbles are $B$-symmetric, but some small fraction are asymmetric. The smallness of the asymmetry in such an example would arise not from small $\epsilon$, but from the exponential suppression of the subdominant instanton.   
Such a model may suffer from fine-tuning issues because the baryon asymmetry would be so sensitive to the Euclidean action of the subdominant instanton.

Finally, in recent years there has been a resurgence of interest in so-called asymmetric dark matter, where the dynamics of baryogenesis and dark matter genesis are linked \cite{otherADM}.  Such a linkage can arise naturally in hidden-sector theories in which dark matter has a $U(1)_\text{DM}$, and can have phenomenological signatures distinct from standard weakly interacting dark matter.   A modification of bubble baryogenesis can achieve simultaneous generation of baryons and dark matter by extending the symmetry structure to $U(1)_B\times U(1)_\text{DM}$.

\begin{acknowledgements}
Thanks to Mustafa Amin, and Paul Steinhardt.  C.~C.~is supported by the Director, Office of Science, Office of High Energy and Nuclear Physics, of the US Department of Energy under Contract DE-AC02-05CH11231, and by the National Science Foundation under grant PHY-0855653.  A.~D.~is supported by the Berkeley Center for Theoretical Physics, department of physics at UC Berkeley and in part by NSF, under contract 32602-13067-44.
\end{acknowledgements}

\end{document}